\documentclass[superscriptaddress,twocolumn,english,floatfix,aps,prd,amsmath,amssymb,longbibliography]{revtex4-2}
\usepackage{epsfig}
\usepackage[latin1]{inputenc}
\usepackage{multirow}
\usepackage{comment}
\usepackage{color}
\usepackage{array}
\usepackage{enumitem}
\usepackage{amsmath}
\usepackage{times}
\usepackage[table,xcdraw]{xcolor}
\usepackage{xcolor}
\usepackage{float}
\usepackage{epsfig}
\usepackage{bm}
\usepackage{subfigure}
\usepackage{amsmath}
\usepackage{float}
\usepackage{amssymb}
\usepackage{placeins}
\usepackage{graphicx}
\DeclareMathOperator{\sech}{sech}
\usepackage[colorlinks=true]{hyperref}
\hypersetup{citecolor=blue,linkcolor=blue,urlcolor=blue}

\begin{document}

\title{A geometrical approach for obtaining exact analytical solutions for a scalar field in a one-dimensional dynamical cavity}

\author{Maxwell F. de Ara\'{u}jo}
\email{maxwell.araujo@icen.ufpa.br}
\affiliation{Faculdade de F\'{i}sica, Universidade Federal do Par\'{a}, 66075-110, Bel\'{e}m, Brazil}

\author{Odemar Juli\~{a}o do N. Neto}
\email{odemar.neto@icen.ufpa.br}
\affiliation{Faculdade de F\'{i}sica, Universidade Federal do Par\'{a}, 66075-110, Bel\'{e}m, Brazil}

\author{Stanley S. Coelho}
\email{stanley.coelho@icen.ufpa.br}
\affiliation{Faculdade de F\'{i}sica, Universidade Federal do Par\'{a}, 66075-110, Bel\'{e}m, Brazil}
\affiliation{Instituto Federal de Educa\c{c}\~{a}o, Ci\^{e}ncia e Tecnologia do Piau\'{i}, Campus S\~{a}o Raimundo Nonato, 64770-000, S\~{a}o Raimundo Nonato, Brazil}

\author{Jeferson Danilo L. Silva}
\email{jdanilo@ufpa.br}
\affiliation{Faculdade de Engenharia da Computa\c{c}\~{a}o e Telecomunica\c{c}\~{o}es, Universidade Federal do Par\'{a}, 66075-110, Bel\'{e}m, Brazil}

\author{Alessandra N. Braga}
\email{alessandrabg@ufpa.br}
\affiliation{Faculdade de F\'{i}sica, Universidade Federal do Par\'{a}, 67130-660, Ananindeua, Brazil}

\author{Danilo T. Alves}
\email{danilo@ufpa.br}
\affiliation{Faculdade de F\'{i}sica, Universidade Federal do Par\'{a}, 66075-110, Bel\'{e}m, Brazil}

\date{\today}
\begin{abstract}
The interaction between quantum fields and moving mirrors can lead to the creation of real particles from vacuum, a phenomenon known as the dynamical Casimir effect.
This effect was first predicted by G. T. Moore in 1969, who considered a real scalar field inside a cavity with a moving mirror in ($1+1$)D, satisfying Dirichlet boundary conditions at the mirrors.
Moore wrote the general solution for the field in terms of a function that satisfies
a functional equation nowadays known as the Moore equation.
In the literature, exact analytical solutions to the Moore equation form a rather small set, restricted to specific classes of laws of motion.
Cole and Schieve [Phys. Rev. A 52, 4405 (1995)] developed a geometrical approach that yielded a general formula for the solution of the Moore equation for cavities initially at rest.
However, its application has always been done through numerical calculations, and their approach has therefore long been regarded solely as a numerical method.
In this work, we show that their method can also be employed as a powerful method for deriving exact analytical solutions, and show the applicability of this approach to deriving analytical solutions to the Moore equation. 
We obtain original exact analytical solutions for an expanding cavity with a uniformly moving mirror,
a contracting cavity with a mirror that approaches the static one asymptotically, and, in addition, we recover the first analytical solution to the Moore equation for a resonant mirror motion, originally obtained by C. K. Law in 1994.
Moreover, while the general formula obtained by Cole and Schieve is expressed in terms of its known trivial solution in a spacetime region where the field was not affected by the mirror motion, here we extend their approach, 
showing that exact solutions can be constructed from any spacetime region where the Moore function 
has already been determined, even by other methods.
We show that this extended geometrical approach allows us to expand the set of known analytical solutions to the Moore equation from the few classes of motion reported in the literature to an infinite set, as it enables the treatment of new classes of laws of motion obtained by combining those presented here with each other as well as with previously known classes of trajectories.
\end{abstract}

\maketitle

\section{Introduction}

The interaction between quantum fields and moving mirrors resulting in the creation of real particles from the quantum vacuum was first investigated by Moore \cite{Moore-1970}, and other pioneering authors in the 1970s  \cite{DeWitt-PhysRep-1975,Fulling-Davies-PRSA-1976,Davies-Fulling-PRSA-1977,Davies-Fulling-PRSA-1977-II,Candelas-JMP-1976,Candelas-PRSA-1977}.
Since then, to this phenomenon, known as the dynamical Casimir effect (DCE), it has been devoted hundreds of articles (see, for instance, Refs. \cite{Dalvit-CasimirPhysics-2011,Dodonov-Phys-2020,Dodonov-Phys-2025} and those therein), with experiments detecting the creation of photons from the vacuum already done successfully \cite{Wilson-Nature-2011,Lahteenmaki-2013, Vezzoli-2019,Schneider-et-al-PRL-2020}.

In his pioneering paper, Moore investigated a real massless scalar field operator $\hat{\phi}$ in a two-dimensional spacetime, satisfying the massless Klein-Gordon equation, inside a cavity formed by a static mirror (located at $x=0$) and a moving mirror with a prescribed law for the movement $x=L(t)$, both imposing Dirichlet boundary conditions to the field, namely $\hat{\phi}(t,0)=0$ and $\hat{\phi}(t,L(t))=0$.
Since the original paper by Moore \cite{Moore-1970}, ($1+1$)D models have had an important role in the investigation of the DCE, having been considered by DeWitt \cite{DeWitt-PhysRep-1975}, Fulling and Davies \cite{Fulling-Davies-PRSA-1976}, and also in many other works (see Refs. \cite{Dodonov-Klimov-PRA-1996,*Alves-Farina-MaiaNeto-JPA-2003,*Alves-Farina-Granhen-PRA-2006,*Alves-Granhen-PRA-2008, *Alves-Granhen-Silva-Lima-PLA-2010, *Alves-Granhen-Silva-Lima-PRD-2010,*Alves-Granhen-CPC-2014,*Good-PRD-2016,*Good-PRD-2020,*Good-Orlando-PRD-2021,*Silva-Farina-PRD-2011,*Silva-Braga-Rego-Alves-PRD-2015,*Silva-Braga-Alves-PRD-2016,*Silva-Braga-Rego-Alves-PRD-2020,*Rego-Braga-Silva-Alves-PRD-2022,Law-PRL-1994,Cole-Schieve-PRA-1995,Cole-Schieve-PRA-2001}). These models also served as part of the theoretical basis for the first experimental observation of photon creation from vacuum, reported in Ref. \cite{Wilson-Nature-2011}.

By exploiting the conformal invariance of the wave equation, $(\partial _{t}^{2}-\partial _{x}^{2})\hat{\phi} \left(t,x\right) =0$ (we assume $\hbar=c=1$ throughout this paper), Moore found that the general form of the solution for the field in such a dynamical cavity is given by \cite{Moore-1970,Fulling-Davies-PRSA-1976},
\begin{equation}
\hat{\phi}(t,x) = \sum_{n=1}^{\infty} \left[
\hat{a}_{n}\,\phi_{n}(t,x) + \hat{a}_{n}^{\dagger}\,\phi_{n}^{*}(t,x)\right],
\label{eq:field-1}
\end{equation} 
where $\phi_{n}(t,x)$ are the field modes given by
\begin{equation}
\phi_{n}(t,x)=\frac{i}{2\sqrt{n\pi}}\left[e^{- i n \pi R(v)}-e^{- i n \pi R(u)}\right], 
\label{eq:field-2}
\end{equation} 
with $u=t-x$, $v=t+x$, and the phase function $R$, known as the Moore function, satisfying the functional equation
\begin{equation}
R[t+L(t)]-R[t-L(t)]=2,
\label{Moore-equation}
\end{equation}
which was also obtained independently by Vesnitskii in Ref. \cite{Vesnitskii-RQE-1971}. 
The function $R$ defines a conformal transformation that maps a given dynamical cavity in the
usual spacetime onto a static cavity in a conformal space \cite{Moore-1970, Fulling-Davies-PRSA-1976}.

Thus, for a given law of motion $L$, the corresponding solution $R$ of the Moore equation determines the dynamics of the confined quantum field and allows the calculation of various physical quantities affected by the mirror's motion
\cite{Fulling-Davies-PRSA-1976,Moore-1970,Alves-Granhen-Alves-Lima-PRD-2020,Dodonov-Phys-2020}.
For example, assuming the vacuum as the initial ($\textit{in}$) state of the field, 
the energy density inside the cavity is determined by the expectation value of the $\hat{T}_{00}$ component of the energy-momentum tensor, which after renormalization is given by
\cite{Moore-1970,Fulling-Davies-PRSA-1976,Cole-Schieve-PRA-2001}
\begin{equation}
	\langle\hat{T}_{00}(t,x)\rangle_{\text{in}}=-f(v)-f(u),
	\label{eq:Tensor-1}
\end{equation}
where
\begin{equation}
	f=\frac{1}{24\pi}\left[\frac{R^{\prime\prime\prime}}{R^{\prime}}-\frac{3}{2}\left(\frac{R^{\prime\prime}}{R^{\prime}}\right)^{2}+\frac{\pi^{2}}{2}R^{\prime2}\right].
	\label{eq:Tensor-2}
\end{equation}
From these formulas the contribution of the confined field to the net force acting on the mirrors can also be obtained \cite{Alves-Farina-MaiaNeto-JPA-2003, Alves-Granhen-Lima-PRD-2008, Cole-Schieve-PRA-2001}.
Moreover, considering the cavity with initial length $L_0$, once the mirror motion ceases, the expectation value ${\cal{N}}$ of the total number of created particles found inside the cavity at a time $t$ in the $\textit{out}$ state is given by \cite{Castagnino-Ferraro-AnnPhys-1984,Alves-Granhen-Alves-Lima-PRD-2020}
\begin{equation}
	{\cal{N}}(t)=\sum_{n,m=1}^{\infty}|\beta_{nm}(t)|^{2},
\end{equation}
where 
\begin{equation}
	\beta_{mn}(t)=-\frac{1}{2}\sqrt{\frac{m}{n}}\int_{t/L_{0}-1}^{t/L_{0}+1}dxe^{-i\pi\left[n R(L_{0}x) + m x\right]},
\end{equation}
form part of the Bogoliubov coefficients. Therefore, the Moore function $R$ plays a central role in describing the dynamics of the system \cite{Moore-1970,Fulling-Davies-PRSA-1976,Cole-Schieve-PRA-2001,Alves-Granhen-Silva-Lima-PRD-2010, Dodonov-Phys-2020, Castagnino-Ferraro-AnnPhys-1984}.

A problem is that there is no formal analytical solution to Eq.~(\ref{Moore-equation}) for a general function $L(t)$,
and the search for analytical solutions to it, and consequently for the quantum field in a dynamical cavity, has been the focus of efforts \cite{Moore-1970, Fulling-Davies-PRSA-1976, Castagnino-Ferraro-AnnPhys-1984, Dodonov-JMP-1993, Law-PRL-1994, Cole-Schieve-PRA-1995,Wu-PRA-1999, Li-Ping-Fu-PRA-1997,Alves-Granhen-CPC-2014},
and remains a topic of current research \cite{Lassuyt-et-al-ZMP-2025}.
Despite more than five decades of interest in this topic, exact analytical solutions to Eq.~(\ref{Moore-equation}) form only a small set, restricted to exceptional laws of motion.
While in the trivial case of a motionless cavity [$L(t)=L_0$] the Moore function is straightforwardly obtained as \cite{Moore-1970}
\begin{equation}
R(z)={z}/{L_0},
\label{R-static}
\end{equation}
non-trivial scenarios demand specific solution methods \cite{Moore-1970, Fulling-Davies-PRSA-1976, Castagnino-Ferraro-AnnPhys-1984,Dodonov-JMP-1993, Law-PRL-1994,  Wu-PRA-1999, Li-Ping-Fu-PRA-1997, Lassuyt-et-al-ZMP-2025}.

\begin{figure}[t]
\epsfig{file=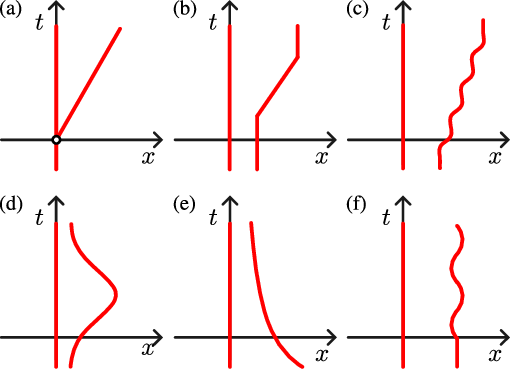,width=1\linewidth}
\caption{
Models of dynamic cavities for which exact solutions to the Moore equation have been obtained in the literature.	
(a) A mirror in uniform motion intersecting the static mirror at the origin \cite{Moore-1970}. 
(b) A static mirror at a given position starts moving uniformly and then stops abruptly at a new position \cite{Moore-1970,Fulling-Davies-PRSA-1976,Castagnino-Ferraro-AnnPhys-1984}. 
(c) A static mirror undergoes resonant oscillation during a time interval and then returns to rest at a new position \cite{Li-Ping-Fu-PRA-1997}. 
(d) A mirror executing a single symmetric pulse that starts and ends at the origin asymptotically \cite{Lassuyt-et-al-ZMP-2025}. 
(e)  A mirror in a gradual deceleration trajectory \cite{Dodonov-JMP-1993}. 
(f) A static mirror starts executing a resonant periodic motion for $t>0$ \cite{Law-PRL-1994,Law1995Tese,Wu-PRA-1999}. 
In red, we denote cavity models existing in the literature.
}
\label{fig:Fig-literature}
\end{figure}

A method for obtaining exact analytical formulas for $R$ was proposed by Moore himself \cite{Moore-1970}, by means of solving an ``inverse problem'', i.e., proposing a given exact analytical formula for $R$ and then finding the corresponding formula for $L(t)$.
With this approach, exact analytical formulas for $R$ were obtained for the following models of cavities:
an initially collapsed ($L_0=0$) at $t=0$, and then with one of the mirrors moving away with a constant velocity \cite{Moore-1970} (Fig. \ref{fig:Fig-literature}a); 
with one of the mirrors that starts at rest, moves uniformly for a finite duration, and returns to rest \cite{Moore-1970} (Fig. \ref{fig:Fig-literature}b);
with one of the mirrors, initially at rest, undergoing a modulated resonant oscillation regime for a finite time interval, and then returning to rest \cite{Li-Ping-Fu-PRA-1997} (Fig. \ref{fig:Fig-literature}c);
with one of the mirrors asymptotically expanding smoothly from the static mirror, reaching a maximum displacement, and asymptotically returning to the static mirror \cite{Lassuyt-et-al-ZMP-2025} (Fig. \ref{fig:Fig-literature}d). 

A static zone is a spacetime region in which the field has not yet been affected by the mirror motion and the solution for $R$ is given by Eq.~\eqref{R-static}.
An initially collapsed cavity, which corresponds to $L_0=0$, has no prior well-defined static zone, since Eq.~\eqref{R-static} is not defined for $L_0=0$.
As previously mentioned, the exact analytical formula for $R$ for this type of cavity [Fig.~\ref{fig:Fig-literature}(a)], undergoing uniform expansion starting at $t=0$, was obtained by Moore via an ``inverse problem'' approach \cite{Moore-1970}.
The exact analytical formula for $R$ for an equivalent model, in which a uniform expansion with velocity $\alpha$ starts at $t = -L_0/\alpha$, was obtained via an appropriate ``reparameterization method'' \cite{Baranov-Shirokov-SPJ-1968, Dodonov-JMP-1993} (see Sec.~\ref{sec:reparamet}).
For a contracting cavity with a mirror that asymptotically approaches a static mirror (Fig.~\ref{fig:Fig-literature}e), the exact analytical solution for $R$ was obtained via a 
perturbative approach that led to a non-zero first term, with all higher-order corrections vanishing \cite{Dodonov-JMP-1993}.
The first exact analytical solution to the Moore equation for a resonant mirror (Fig. \ref{fig:Fig-literature}f) motion was obtained in Refs. \cite{Law-PRL-1994,Law1995Tese}, by exploiting the periodicity of the motion and mapping the continuous Moore equation onto a discrete recurrence relation.

The difficulty of finding exact analytical solutions to Eq. (\ref{Moore-equation}) led several authors
(including Moore) to seek approximate solutions via perturbative methods \cite{Moore-1970,Dodonov-JMP-1993,Dalvit-Mazzitelli-PRA-1999}
and asymptotic expansions \cite{Dodonov-Klimov-PLA-1992}.
It is worth mentioning that several authors \cite{Ford-Vilenkin-PRD-1982}, although they did not directly investigate perturbative analytical solutions to Eq.~(\ref{Moore-equation}), investigated perturbative solutions for the field operator $\hat{\phi}(t,x)$ \cite{Ford-Vilenkin-PRD-1982,Razavy-PRD-1985,Dodonov-Klimov-PRA-1996}. 
Numerical perturbative solutions for $\hat{\phi}(t,x)$ have also been investigated
\cite{Ruser-JOptB-2005,Ruser-JPA-2006}.

Another very interesting approach is the exact geometric method proposed by Cole and Schieve \cite{Cole-Schieve-PRA-1995}, which yielded a general formula [Eq. \eqref{R(z)-Geometric}] for solving Eq.~\eqref{Moore-equation}, for cavities initially at rest.
However, its application has always been done through numerical	recursive calculations, and their approach has therefore long been regarded solely as a numerical method for solving the Moore equation \cite{Cole-Schieve-PRA-1995, Alves-Granhen-Lima-Rego-JPCS-2009, Alves-Granhen-PRA-2008, Alves-Granhen-Lima-PRD-2008, Alves-Granhen-Silva-Lima-PLA-2010, Alves-Granhen-Silva-Lima-PRD-2010, Alves-Granhen-CPC-2014, Wu-PRA-1999, Xue-Ying-CTP-2001, Lassuyt-et-al-ZMP-2025, Villar-Mazzitelli-Entropy-2023,BoZangLi-PLA-2002}.

\begin{figure}[t]
	\epsfig{file=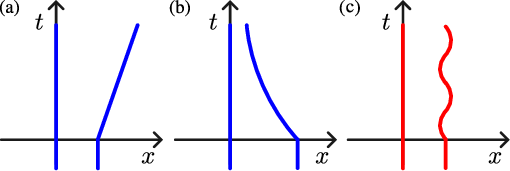, width=1\linewidth}
	\caption{The three cavity models investigated here, for which exact solutions to the Moore equation are obtained using the analytical geometric method. In blue, we denote the new cavity models introduced here, whereas in red we denote an existing one from the literature. In each configuration, an initially static mirror starts executing a specific law of motion for $t>0$: (a) Uniform motion. (b) Gently contracting motion. (c) Resonant periodic motion \cite{Law-PRL-1994}.}
	\label{fig:Nossos-movimentos}
\end{figure}

In this work, we present a novel method for deriving exact analytical solutions to the Moore equation. 
We demonstrate that the geometric approach developed by Cole and Schieve \cite{Cole-Schieve-PRA-1995} 
can also be employed as a powerful method for deriving exact analytical solutions.
By applying this method, we obtain novel exact analytical solutions for two classes of cavity motion in which an initially static mirror is set into motion: an expanding cavity with a uniformly moving mirror (Fig.~\ref{fig:Nossos-movimentos}(a)) and a contracting cavity with a mirror that asymptotically approaches the static one (Fig.~\ref{fig:Nossos-movimentos}(b)).
%
%
%
In addition, we recover the first exact analytical solution to the Moore equation for a resonant mirror motion (Fig.~\ref{fig:Nossos-movimentos}c), originally obtained by Law \cite{Law-PRL-1994}.
Note that Figs.~\ref{fig:Nossos-movimentos}(a) and \ref{fig:Nossos-movimentos}(b) are shown in blue, whereas Fig.~\ref{fig:Nossos-movimentos}(c) is shown in red.
The color convention adopted throughout this work uses blue to denote new cavity models [new laws of motion $L(t)$] introduced here, for which we derive original exact solutions to the Moore equation, and red to identifies cavity models existing in the literature for which exact solutions are known.

\begin{figure}[t]
	\epsfig{file=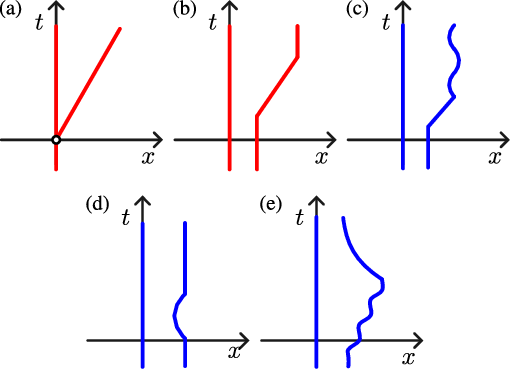, width=1\linewidth}
	\caption{The five cavity models investigated here, for which exact solutions to the Moore equation have been obtained via an extension of the geometric method. In blue, we denote the new cavity models introduced here, whereas in red we denote those existing in the literature. (a)  A mirror in uniform motion intersecting the static mirror at the origin \cite{Moore-1970}. (b)  A static mirror at a given position starts moving uniformly and then stops abruptly at a new position \cite{Moore-1970, Castagnino-Ferraro-AnnPhys-1984, Fulling-Davies-PRSA-1976}. (c) A static mirror at a given position starts moving uniformly before transitioning into resonant motion. (d) A mirror in a single oscillation semi-cycle with an abrupt halt. (e) A mirror in modulated resonant oscillation, with a subsequent asymptotic contraction.}
	\label{fig:3-intro-extension}
\end{figure}
\begin{figure}[t]
	\centering
	\epsfig{file=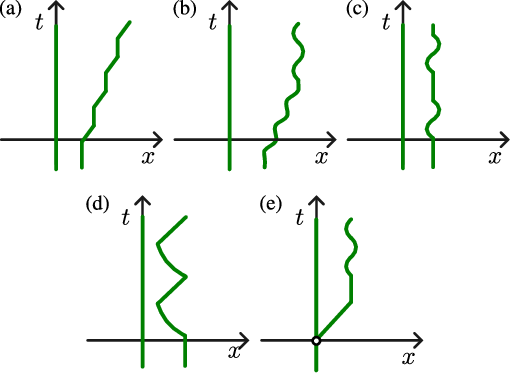, width=1\linewidth}
	\caption{Some of the infinitely many other models for which 
		analytical exact solutions to the Moore equation may be obtained via the extension of the geometric method proposed here (the green color distinguishes these models from those with explicitly obtained analytical exact solutions). (a) A mirror executing an alternating sequence of uniform motion and abrupt halts. (b) A mirror undergoing a modulated resonant oscillation, followed by a transition to a distinct resonant periodic regime. (c) A mirror executing a single oscillation cycle, halting abruptly, and then undergoing a second single oscillation cycle.  (d) A mirror executing an alternating sequence of uniform expansion and asymptotic contraction  (e) A mirror, initially collapsed with the static one, executing a uniform motion, followed by an abrupt halt and the onset of resonant periodic motion.}
	\label{fig:4-intro-extension}
\end{figure}

Furthermore, while the general formula obtained by Cole and Schieve \cite{Cole-Schieve-PRA-1995} is expressed in terms of the known static-zone solution [Eq.~\eqref{R-static}], here we extend their approach in two ways.
First, we enable the geometric method to describe initially collapsed cavities ($L_0=0$), which possess no prior well-defined static zone.
By doing so, we recover the analytical solutions for initially collapsed mirrors moving uniformly, encompassing both Moore's model \cite{Moore-1970} (Fig. \ref{fig:3-intro-extension}a) and that investigated by Dodonov, Klimov and Nikonov, in which the initial cavity collapse occurs at $t = -L_0/\alpha$ \cite{Dodonov-JMP-1993}.
Second, we demonstrate that new exact solutions can be constructed from any spacetime region where the function $R(z)$ 
has already been previously determined, even by other methods.
With this latter extension, we expand the set of known analytical solutions to the Moore equation from a small set of models found in the literature to an infinite one, as it allows us to handle new laws of motion by combining those considered here with one another and with those found in other works.
To illustrate this, we derive exact analytical solutions for four novel cases formed by combining different motions: a uniform expansion ending in an abrupt halt (Fig.~\ref{fig:3-intro-extension}b), a uniform expansion transitioning into a resonant motion (Fig.~\ref{fig:3-intro-extension}c), a single oscillation semicycle with an abrupt halt (Fig.~\ref{fig:3-intro-extension}d), and a modulated resonant oscillation, with a subsequent asymptotic contraction (Fig.~\ref{fig:3-intro-extension}e).
In addition, Fig.~\ref{fig:4-intro-extension} illustrates some of the infinitely many other models for which analytical exact solutions to the Moore equation may be obtained via the extension of the geometric method proposed here (the green color is used to visually distinguish these models from the others for which analytical exact solutions were obtained explicitly). 

The first model we investigate (Fig.~\ref{fig:Nossos-movimentos}a) has the advantage of
yielding a finite static Casimir force before the mirror motion begins (as expected), whereas in the collapsed-cavity model found in the literature \cite{Moore-1970} (Fig.~\ref{fig:Fig-literature}a) the Casimir force diverges before the motion begins.
Our second model (Fig.~\ref{fig:Nossos-movimentos}b) has advantage when
compared to the literature \cite{Dodonov-JMP-1993} (Fig.~\ref{fig:Fig-literature}e)
because, by setting $L_{0}\alpha < 1$, our model departs from a static configuration with a subluminal velocity, whereas the model illustrated in Fig.~\ref{fig:Fig-literature}e involves superluminal velocities at early times.
In addition, by investigating the model shown in Fig.~\ref{fig:Fig-literature}b, we obtain a general analytical solution to the corresponding Moore equation, written in terms of four degrees of freedom, given by the initial length, start time, stop time, and expansion velocity. 
On the other hand, the solution found in the literature \cite{Moore-1970} 
is less general, since it is written in terms of only three degrees of freedom,
which means that there is an intrinsic constraint between the mentioned parameters.
In this way, our solution offers an advantage over the model in the literature \cite{Moore-1970}.

The paper is organized as follows. In Sec.~\ref{sec:Literature-Overview}, we provide an overview of exact analytical methods found in the literature.
In Sec.~\ref{Geometric-Method}, we provide an overview of the geometric method \cite{Cole-Schieve-PRA-1995}.
Using this method, we obtain original exact analytical solutions for two cases:  in Sec.~\ref{sec:Uniform-Motion}, an expansion with uniform motion; 
in Sec.~\ref{sec:gentle_contraction}, a gentle contraction. 
In Sec. \ref{sec:resonant_motion}, we recover, via geometric method, the first analytical solution to the Moore equation for a resonant mirror motion \cite{Law-PRL-1994}.
In Sec.~\ref{sec:extension-Geometric-Method-Singularity-1}, we present a first extension of the geometric method, enabling it to encompass initially collapsed cavities.
In Sec.~\ref{sec: extension-Geometric-Method-2}, we present a second extension, enabling the geometric method to yield exact analytical solutions involving combined laws of motion.
Sec.~\ref{final-comments} is devoted to our final comments.

\section{An Overview of Exact Analytical Methods in the Literature}
\label{sec:Literature-Overview}
This section provides an overview of the analytical methods and moving boundary models corresponding to exact solutions to Moore equation found in the literature \cite{Moore-1970, Dodonov-JMP-1993, Law-PRL-1994, Dodonov1996Book,Law1995Tese,Baranov-Shirokov-SPJ-1968, Fulling-Davies-PRSA-1976, Castagnino-Ferraro-AnnPhys-1984, Wu-PRA-1999, Li-Ping-Fu-PRA-1997, Lassuyt-et-al-ZMP-2025}. We shall review the approch via inverse problem introduced by Moore \cite{Moore-1970}, the ``reparameterization method''  \cite{Baranov-Shirokov-SPJ-1968,Dodonov-JMP-1993}, the perturbative method \cite{Dodonov-JMP-1993, Moore-1970},  and finally the method developed by Law \cite{Law-PRL-1994,Law1995Tese}.

\subsection{The Inverse Problem Method}
\label{subsec:inverse-problem}
To circumvent the intrinsic difficulties of solving Eq.~(\ref{Moore-equation}) directly, Moore \cite{Moore-1970} pointed out that one way to obtain analytical exact formulas for $R$ is by means of an ``inverse problem'', i.e., proposing a given exact analytical formula for $R$ and then finding the corresponding formula for $L(t)$, resulting in a purely algebraic problem.
For instance, the case  $L(t)=\alpha t$ for $t > 0$, analyzed by Moore \cite{Moore-1970}, describes a cavity initially collapsed and then one of the mirrors moves in uniform motion from the static mirror at the origin (Fig. \ref{fig:Fig-literature}a). Using this inverse problem method, Moore obtained the exact analytical solution given by
\begin{equation}
	R(z)=\frac{2\ln z}{\ln\left[(1+\alpha)/(1-\alpha)\right]}.
	\label{eq:Moore-solution}
\end{equation}

Moore \cite{Moore-1970} also considered the case of a mirror that starts at rest, moves uniformly for a finite duration, and returns to rest (Fig. \ref{fig:Fig-literature}b). This law of motion is described as follows:
\begin{equation}
	L(t)=\begin{cases}
		\tau/\alpha_{1}, & t<\tau-\tau/\alpha_{1},\\
		\frac{(\alpha_{1}-\alpha_{2})(t-\tau)+2\tau}{\alpha_{1}+\alpha_{2}},& \tau-\tau/\alpha_{1}<t<\tau+\tau/\alpha_{2},\\
		\tau/\alpha_{2}, & t>\tau+\tau/\alpha_{2},
		\label{eq:Moore-lei-MRU}
	\end{cases}
\end{equation}
where the parameters $\alpha_1$, $\alpha_2$, and $\tau$ adjust the profile of the moving mirror's worldline. 
Notice that the stopping point and the speed of the mirror are constrained, and the mirror is required to stop somewhere; otherwise, it will reach the speed of light. 
For this trajectory, the exact analytical solution obtained for the Moore equation is given by \cite{Moore-1970}
\begin{equation}
	R(z)=\begin{cases}
		\alpha_{1}(z-\tau)/\tau, & z<\tau,\\
		\alpha_{2}(z-\tau)/\tau, & z>\tau.
	\end{cases}
	\label{eq:R(z)-Moore-MRU}
\end{equation}
This model was revisited by Fulling and Davies, in Ref. \cite{Fulling-Davies-PRSA-1976}, where they analyzed the energy density of the field inside the corresponding dynamical cavity. Similar systems were also investigated by Castagnino and Ferraro \cite{Castagnino-Ferraro-AnnPhys-1984}, focusing on the calculation of Bogoliubov coefficients to quantify particle creation.

For the model illustrated in Fig. \ref{fig:Fig-literature}c, Fu, Duan and Guo \cite{Li-Ping-Fu-PRA-1997} found, via the inverse problem method, an analytical exact formula for $R$ given by
\begin{equation}
	L(t) = \begin{cases} 
		1/(1+\lambda) , & t^{\prime} \le 0, \\
		\dfrac{1+\lambda t^{\prime}}{1+\lambda} + Q(t^{\prime}), & 0 \le t^{\prime} \le 1, \\
		1, & t^{\prime} \ge 1,
	\end{cases}
	\label{eq:L(t)-Fu}
\end{equation}
where $t=[\left(2+\lambda\right)t^{\prime}-1]/(1+\lambda)$, 
and 
\begin{equation}
	Q(t^{\prime})=-\frac{\lambda}{m\pi(1+\lambda)}\sin\left[2m\pi t^{\prime}+m\pi Q(t^{\prime})\right],
	\label{eq:Q(t)-Fu}
\end{equation}
in which the positive integer parameter $m$ dictates the frequency of oscillation and the amplitude of the motion, and $|\lambda| < 1$, to avoid superluminal speeds. Under these conditions, the exact analytical solution to the Moore equation is given by \cite{Li-Ping-Fu-PRA-1997}
\begin{equation}
	R(z) = \begin{cases}
		(2+\lambda)z, & z \le 0, \\
		z + \dfrac{\lambda}{m\pi}\sin(m\pi z), & z > 0.
	\end{cases}
	\label{eq:R(z)-Fu}
\end{equation}
It is worth noting that, in the special case $m = 1$, the model given by Eqs.~(\ref{eq:L(t)-Fu})--(\ref{eq:Q(t)-Fu}) reproduces a specific model investigated in Ref.~\cite{Castagnino-Ferraro-AnnPhys-1984}.

More recently, the inverse problem method was applied by Lassuyt et al. \cite{Lassuyt-et-al-ZMP-2025} for a model where one of the mirrors describes a single symmetric pulse motion, expanding smoothly from the static mirror, and asymptotically returning to the static mirror, as illustrated in Fig. \ref{fig:Fig-literature}d. 
This trajectory is described by
\begin{equation}
	L(t) = \frac{1}{k} \arcsin \left\{ \frac{1}{A} \sech \left[ k(t-z_{0}) \right] \right\},
	\label{eq:L(t)-Lassuyte}
\end{equation}
where the constant $z_0$ determines the temporal center of the pulse, $k$ controls the amplitude of the motion, and $A$ is a dimensionless constant.
The exact analytical solution to the Moore equation is
\begin{equation}
	R(z)=A\sinh\left[k(z-z_{0})\right].
	\label{eq:R(z)-Lassuyte}
\end{equation}

To our knowledge, these are all the exact analytical solutions found in the literature and obtained via the inverse problem method.
%
\subsection{The Reparameterization Method}
\label{sec:reparamet}

An alternative approach to obtain analytical exact formulas for $R$ is through a ``reparameterization method'' introduced by Dodonov, Klimov and Nikonov \cite{Dodonov-JMP-1993}, based on the calculations by Baranov and Shirokov \cite{Baranov-Shirokov-SPJ-1968}.
This method consists of introducing a specific change of variable that maps the boundary condition at the moving mirror into a periodic functional equation.
By expanding the general solution as a Fourier series in the new variables, the exact analytical form of $R$ can be directly extracted for a given trajectory $L(t)$.

By applying this method to solve the Moore equation, the authors in Ref.~\cite{Dodonov-JMP-1993} analyzed a uniformly moving mirror given by the trajectory 
\begin{equation}
	L(t) = L_{0} + \alpha t,
\end{equation}
valid for $t > -L_{0}/\alpha$.
This model is equivalent to the uniformly expanding cavity originally analyzed by Moore \cite{Moore-1970} (Fig.~\ref{fig:Fig-literature}a), but with the cavity collapse shifted from $t=0$ to $t = -L_{0}/\alpha$.
Under these conditions, the exact analytical solution is given by \cite{Dodonov-JMP-1993}
\begin{equation}
	R(z)=\frac{2\ln\left(1+\alpha z/L_{0}\right)}{\ln\left[(1+\alpha)/(1-\alpha)\right]}.
	\label{eq:Dodonov/Baranov-solution}
\end{equation}
%
\subsection{The Perturbative Method}

An interesting particular scenario in which a perturbative approach \cite{Dodonov-JMP-1993,Moore-1970} leads to an exact result was analyzed by Dodonov, Klimov and Nikonov \cite{Dodonov-JMP-1993}. 
These authors considered a contracting cavity with a mirror that asymptotically approaches a static mirror (as illustrated in Fig. \ref{fig:Fig-literature}e), following the trajectory 
\begin{equation}
	L(t)=L_{0}/(1+\alpha t),
\end{equation}
where $t>-1/\alpha$.
Surprisingly, for this specific law of motion, the perturbative solution is exact since all higher-order  corrections vanish, leaving only the zeroth-order term given by \cite{Dodonov-JMP-1993}
\begin{equation}
	R(z)=(z+\alpha z^{2}/2)/L_{0}.
	\label{eq:dodonov_solution}
\end{equation}

\subsection{Law's Method}

Law \cite{Law-PRL-1994,Law1995Tese} developed a method to find the first exact analytical solution for a resonant moving cavity.
This solution was particularly important for understanding the DCE,
since with it Law showed that the finite part of the field energy density in the cavity develops into two traveling wave packets that become narrower as time increases, and that a ``sub-Casimir'' region emerges, where the energy density can be even lower than the usual static Casimir energy density \cite{Law-PRL-1994}.
In addition, this model is relevant because resonant motions can significantly increase 
the field force on the mirrors \cite{Jaekel-Reynaud-PLA-1992,Dodonov-JMP-1993}, and
can also lead to an exponential growth in the number of created particles \cite{Dodonov-Klimov-PRA-1996}. 
 
According to Ref.~\cite{Law1995Tese}, the method
assumes $L(t)=L_0-\epsilon(t)$, where $\epsilon(t\leq 0)=0$ and $\epsilon(t)=\epsilon(t+2L_0)$ for $t>0$ [$\epsilon(t)$ can be expanded in the cavity's eigenfrequencies $n\pi/L_0$, where $n$ is a non-negative integer]. 
Writing $R(z)=z/L_0+h(z)/L_0$ and substituting this into the Moore equation yields $h(t+L_0-\epsilon(t))=h(t-L_0+\epsilon(t))+2\epsilon(t)$. 
To solve this, Law introduces a variable $\xi\in[-L_{0},L_{0})$ such that $t+L_{0}-\epsilon(t)=2nL_{0}+\xi$. By defining $h_{n}(\xi)\equiv h(2nL_{0}+\xi)$ and $\phi_{n}(\xi)=h_{n}(\xi)+\xi$, the problem reduces to the recurrence relation $\phi_{n+1}(\xi)\equiv\phi_{n}(\xi)+w\left(\phi_{n}(\xi)\right)$, where $w \equiv 2\epsilon(t)$, and the function $R$ is given by $R(2nL_{0}+\xi)=2n+{\phi_{n}(\xi)}/{L_{0}}$.
In the small-amplitude and nonrelativistic regimes, the recurrence relation between $\phi_{n+1}(\xi)$ and $\phi_{n}(\xi)$ can be solved, providing an approximate analytical solution for $\phi_{n}(\xi)$ and, consequently, for $R$.
As a concrete example, Law considered the motion $\epsilon(t)=\mu\sin^{2}\left({2\pi t}/{L_{0}}\right)$, where the oscillation amplitude $\mu$ is small compared to $L_{0}$, and obtained the approximate function $R$ given by
\begin{equation} 
	R(2nL_{0}+\xi)=2n+\frac{1}{2}-\frac{1}{\pi}\arctan\left(\cot\frac{\xi\pi}{L_{0}}-\frac{2n\epsilon\pi}{L_{0}}\right).
	\label{eq:Law_solution}
\end{equation}
Assuming that this approximate solution could be exact for another law of motion, 
Law used the inverse problem method (Sec. \ref{subsec:inverse-problem}) and, substituting Eq.~(\ref{eq:Law_solution}) into Eq.~(\ref{Moore-equation}), obtained the following resonant periodic law of motion (see Fig.~\ref{fig:Fig-literature}f):
\begin{equation}
	L(t)=L_{0}\left[1+\frac{1}{2\pi}\arcsin\left(\sin\theta\cos\frac{2\pi t}{L_{0}}\right)-\frac{\theta}{2\pi}\right],
	\label{eq:Law_motion}
\end{equation}
where $\theta = \arctan(\mu\pi/L_{0})$. 
In other words, for the law of motion defined by Eq.~(\ref{eq:Law_motion}), the solution for $R$ is given by Eq.~(\ref{eq:Law_solution}), and this 
was the first exact analytical solution to the Moore equation for a resonant mirror motion.
An extension of this model, yielding a family of exact solutions for higher-order resonant couplings, was presented in Law's doctoral thesis in 1995 \cite{Law1995Tese}.
Subsequently, this same class of solutions was derived by Wu et al. in 1999 \cite{Wu-PRA-1999}.
%
%
\section{The Geometric Method}
\label{Geometric-Method}

Cole and Schieve \cite{Cole-Schieve-PRA-1995} proposed a geometrical method for solving the Moore equation in cavities initially at rest. A brief discussion of this method is given next.
\begin{figure}
	\epsfig{file=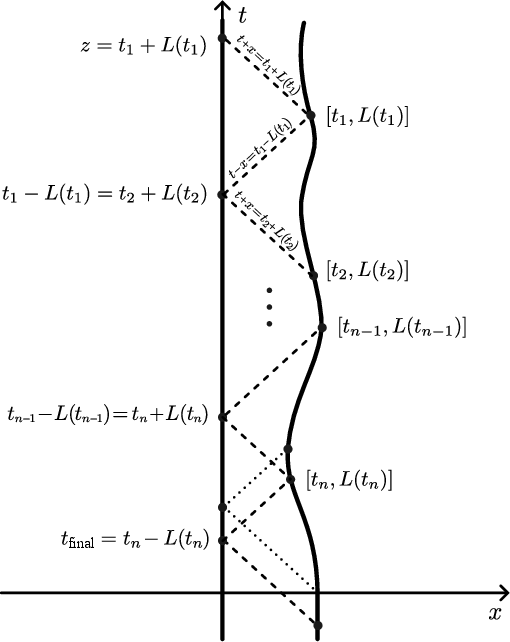,angle=00,width=1.0 \linewidth,clip=}
	\caption{The solid lines represent the trajectories of the cavity boundaries, where the left mirror remains static at $x=0$, while the right mirror remains at rest for $t \leq 0$ before executing an arbitrary law of motion for $t>0$. The dashed straight lines correspond to the trajectories of null lines along the moving cavity.}
	\label{Linhas-nulas}
\end{figure}

Let us consider the null line 
\begin{equation}
v = t_{1} + L(t_{1}) = z,
\label{t1(z)}
\end{equation}
which intersects the moving mirror trajectory at the point $[t_1,L(t_1)]$ (Fig.~\ref{Linhas-nulas}). Using Eq. \eqref{Moore-equation}, one has $R(z)=R[t_1-L(t_1)]+2$.
If $t_1-L(t_1)>L_0$ (as illustrated in Fig.~\ref{Linhas-nulas}), 
one can draw another null line $u=t_1-L(t_1)$, intersecting the world line 
of the static boundary at the point $[t_1-L(t_1),0]$. 
Assuming that the null line $v=t_2+L(t_2)$
intersects the world line of the moving boundary at the point $[t_2,L(t_2)]$ (Fig.~\ref{Linhas-nulas}), we get $t_2+L(t_2)=t_1-L(t_1)$, and using Eq. \eqref{Moore-equation} one can write $R(z)=R[t_2+L(t_2)]+2$. 
By considering Eq. \eqref{Moore-equation}, one can write 
$R(z)=R[t_2-L(t_2)]+4$. 
According to Cole and Schieve, ``the value of $R$ increases by 2 every time there is a reflection off the moving wall'' \cite{Cole-Schieve-PRA-1995}. 
Repeating this process, one obtains a sequence of temporal recurrence equations given by
\begin{equation}
t_{i+1} + L(t_{i+1}) \;=\; t_{i} - L(t_{i}).
\label{recorrencia-ti}
\end{equation}
Then, one traces back a sequence of null lines until one of them enters the static zone  ($u\leq L_{0}$), where the Moore equation reduces to $R[t+L_0]-R[t-L_0]=2$, and the function $R$ is given by Eq.~(\ref{R-static}).
By considering $n(z)$ the number of reflections from the moving mirror
to connect the null line $v=z$ to a null line in the static zone, one obtains
\begin{equation}
R(z) = 2n(z) + R[t_{n}(z)-L(t_{n}(z))].
\label{Eq-6}
\end{equation}
Finally, using Eq.~(\ref{R-static}) into Eq. (\ref{Eq-6}), Cole and Schieve obtained a general formula for the solution of the Moore equation for cavities initially at rest, given by \cite{Cole-Schieve-PRA-1995}
\begin{equation}
R(z) = 2n(z) + [t_{n}(z) - L(t_{n}(z))]/L_{0},
\label{R(z)-Geometric}
\end{equation}
where $t_{n}(z)$ represents the instant of the final reflection on the moving mirror before the null line reaches the static region.

The application of Eq.~(\ref{R(z)-Geometric}) has always been done through numerical calculations, leading the geometric method to be regarded, for more than three decades, solely as a numerical method for determining $R(z)$ \cite{Cole-Schieve-PRA-1995, Alves-Granhen-Lima-Rego-JPCS-2009, Alves-Granhen-PRA-2008, Alves-Granhen-Lima-PRD-2008, Alves-Granhen-Silva-Lima-PLA-2010, Alves-Granhen-Silva-Lima-PRD-2010, Alves-Granhen-CPC-2014, Wu-PRA-1999, Xue-Ying-CTP-2001, Lassuyt-et-al-ZMP-2025, Villar-Mazzitelli-Entropy-2023, BoZangLi-PLA-2002}.
We can see two reasons for this long-standing view.
First, when the laws of motion $L(t)$ are chosen in such a way that the recurrence relations given by Eqs.~(\ref{t1(z)}) and (\ref{recorrencia-ti}) are transcendental equations, there is no analytical solutions for $t_i(z)$ and, consequently, for $n(z)$ and $R(z)$.
Second, even for laws of motion $L(t)$ for which Eqs. \eqref{t1(z)} and \eqref{recorrencia-ti} can be analytically inverted to obtain $t_i(z)$, to the best of our knowledge, no article has addressed calculations to obtain analytical formulas for $n(z)$.
In contrast, we address these calculations next.

\section{Application 1: An Uniformly Expanding Cavity}
\label{sec:Uniform-Motion}
\begin{figure}[t]
\epsfig{file=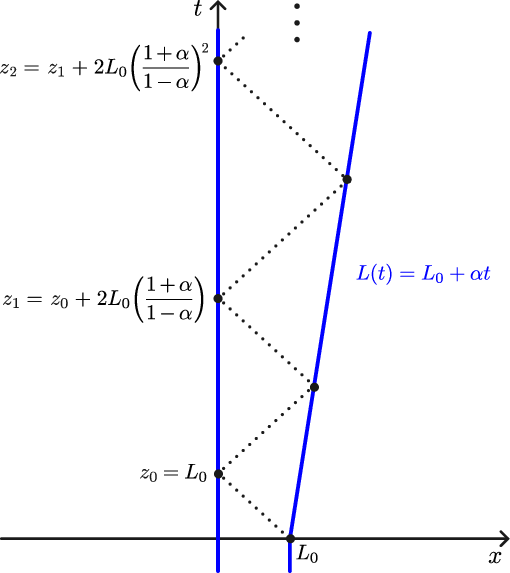,angle=00,width=1.0 \linewidth,clip=}
\caption{Evolution of the expanding cavity. The solid blue lines represent the trajectories of the cavity boundaries, where
the left mirror remains at $x=0$, while the other one moves away at a constant velocity $\alpha$ from rest at $x=L_0$. The dotted black lines represent null trajectories establishing the boundaries of the regions used for counting the number of reflections $n(z)$ inside the cavity required to reach the static zone.
}
	\label{Linhas-nulas-MRU}
\end{figure}

In this section, we outline the way for deriving exact analytical solutions to the Moore equation using the geometric method \cite{Cole-Schieve-PRA-1995}.
To illustrate this, we apply the geometric method to a mirror that is initially static at $x=L_{0}$ for $t\leq0$, and then starts moving uniformly [see Fig. \ref{fig:Nossos-movimentos}a, and more details in Fig. \ref{Linhas-nulas-MRU}], as described by the law of motion:
\begin{equation}
L(t) = L_{0} + \alpha t.
\label{L(t)-MRU}
\end{equation}
The first step in obtaining the exact analytical solution is to apply Eq.~(\ref{L(t)-MRU}) into Eqs.~(\ref{t1(z)}) and~(\ref{recorrencia-ti}). After solving the temporal recurrence for the $n$-th reflection, we obtain
\begin{equation}
t_{n}(z)=\frac{\alpha z+L_{0}}{\alpha(1-\alpha)}\left(\frac{1-\alpha}{1+\alpha}\right)^{n(z)}-\frac{L_{0}}{\alpha}.
\label{tn(z)}
\end{equation}
Substituting Eqs.~(\ref{L(t)-MRU}) and (\ref{tn(z)}) into Eq.~(\ref{R(z)-Geometric}), we obtain
\begin{equation}
R(z)=2n(z)+\frac{\alpha z+L_{0}}{\alpha L_{0}}\left(\frac{1-\alpha}{1+\alpha}\right)^{n(z)}-\frac{1}{\alpha}.
\label{eq:Rz_inicial}
\end{equation} 
The next step is to determine $n(z)$.

For a general geometric view, in Fig.~\ref{Linhas-nulas-MRU} we trace the null lines corresponding to this motion, along with the space-time evolution of the expanding cavity.
The dotted black null lines delimit the regions associated with the number reflections required for a generic null trajetory to reach the static region: if it starts from the range $v=z \leq L_{0}$, it is already in the static region and does not undergo reflection at the moving boundary; if $L_{0} < z \leq L_{0} + 2L_{0}(1+\alpha)/(1-\alpha)$, one reflection is necessary to enter the static region. Taking into account the other regions, we can write
\begin{equation}
n(z)=\sum_{j=0}^{\infty}\Theta(z-z_{j}),
\label{eq:n_z_steps}
\end{equation}
where $\Theta$ is the Heaviside step function, $z_{0}=L_{0}$ and 
\begin{equation}
z_{j}=L_{0}+2L_{0}\sum_{k=1}^{j}\left(\frac{1+\alpha}{1-\alpha}\right)^{k}=\frac{L_{0}}{\alpha}\left[\frac{(1+\alpha)^{j+1}}{(1-\alpha)^{j}}-1\right],
\label{Sn(z)}
\end{equation}
for $j\ge1$, since the sum in $z_{j}$ is the geometric series obtained recursively from the pattern shown in Fig. \ref{Linhas-nulas-MRU}.

The number of reflections, $n(z)$, will be determined from the relation
\begin{equation}
z_{n-1}<z\leq z_{n},
\label{Sn(z)_}
\end{equation} 
representing the interval for which the number of reflections on the moving mirror is $n$ for any $z$. 
%
By substituting Eq. (\ref{Sn(z)}) into (\ref{Sn(z)_}) and performing an algebraic inversion, we obtain
\begin{equation}
n-1<\frac{\ln[(\alpha z/L_{0}+1)/(1+\alpha)]}{\ln[(1+\alpha)/(1-\alpha)]}\leq n.
\label{21}
\end{equation}
This inequality is identified as the definition of the ceiling function \cite{Bender1991}, 
\begin{equation}
	\lceil y \rceil = n \iff n - 1 < y \leq n.
	\label{eq:ceiling_def}
\end{equation}
To satisfy the physical requirement that no reflections occur in the static region where $z \leq L_0$, we must set it to zero for $z \leq L_0$. Therefore, we write $n(z)$ as
\begin{equation}
	n(z)=
		\left\lceil \frac{\ln[(\alpha z/L_{0}+1)/(1+\alpha)]}{\ln[(1+\alpha)/(1-\alpha)]}\right\rceil \text{ for } z\ge L_{0},
	\label{n(z)-ceil}
\end{equation}
and $n(z)=0$ for $z<L_0$.

Finally, by combining Eqs.~(\ref{eq:Rz_inicial}) and (\ref{n(z)-ceil}), we obtain the final exact analytical expression for $R(z)$. 
To ensure the correctness of this solution, we subjected it to some consistency tests. First, it satisfies Moore's functional equation [Eq. \eqref{Moore-equation}]. Moreover, in the limit $\alpha \rightarrow 0$, it recovers the solution of the static case [given by Eq. (\ref{R-static})]. Lastly, our analytical solution [Eqs. (\ref{eq:Rz_inicial}) and (\ref{n(z)-ceil})] is in agreement with the results obtained numerically via the Dynamic Casimir Package (DCP) \cite{Alves-Granhen-CPC-2014}.
%

\section{Application 2: An Asymptotic Contracting Cavity}\label{sec:gentle_contraction}

In this section, we apply the geometric method to a asymptotic contracting cavity, whose moving mirror follows the law of motion
 \begin{equation}
 	L(t) = \frac{L_{0}}{1 + \alpha t},
 	\label{eq:L-asymptotic}
 \end{equation}
for $t \ge 0$. For $t<0$, the mirror remains at rest at $x=L_0$. 
Following the same steps of the previous section, we first apply Eq.~(\ref{eq:L-asymptotic}) into Eqs.~(\ref{t1(z)}) and~(\ref{recorrencia-ti}).
Then, solving the recurrence relation, we obtain 
\begin{equation}
t_{n}=\frac{\sqrt{\kappa-4\alpha L_{0}(n-2)}+\sqrt{\kappa-4\alpha L_{0}(n-1)}-2}{2\alpha},\label{eq:tn_asymptotic}
\end{equation}
where, for the sake of simplicity, we define the dimensionless parameter $\kappa=[1+\alpha t_{1}-\alpha L_{0}/(1+\alpha t_{1})]^{2}$, with
\begin{equation}
t_{1}=\frac{\alpha z-1+\sqrt{(1+\alpha z)^{2}-4\alpha L_{0}}}{2\alpha}.
\label{eq:t1_asymptotic}
\end{equation}
Then, substituting Eqs.~(\ref{eq:tn_asymptotic}) and~(\ref{eq:t1_asymptotic}) into Eq.~(\ref{R(z)-Geometric}), we obtain
\begin{equation}
R(z)=2n(z)+\frac{1}{\alpha}\left[\sqrt{(1+\alpha z)^{2}-4\alpha L_{0}n(z)}-1\right],
\label{eq:Rz_asymptotic}
\end{equation}
where $n(z)$ corresponds to the exact number of reflections off the moving mirror required to reach the static zone.

\begin{figure}
\epsfig{file=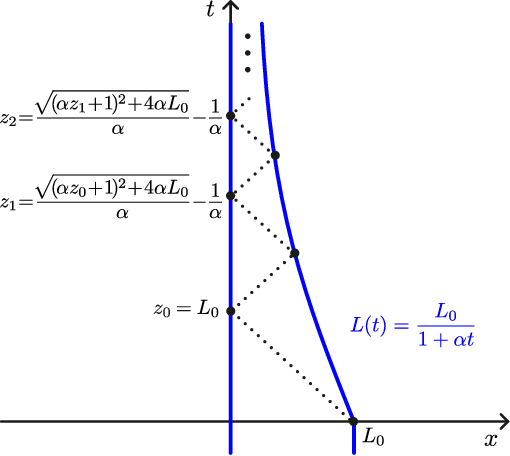,angle=00,width=1.0 \linewidth,clip=}
\caption{
Evolution of the gently contracting cavity. 
The solid blue lines represent the trajectories of the cavity boundaries, where the left mirror remains static at $x=0$, while the right one gradually approaches the static mirror asymptotically, governed by Eq. (\ref{eq:L-asymptotic}). The dotted lines represent null trajectories establishing the boundaries of the regions used for counting the number of reflections, $n(z)$ inside the cavity until reaching the static zone.
}
\label{Diagrama-Suave}
\end{figure}

In order to count the number of reflections $n(z)$, we consider the spacetime diagram and the associated null trajectories shown in
Fig.~\ref{Diagrama-Suave}. Each interval $v\in(z_{j-1}, z_{j})$ [with $(j=1,2, \dots)$] corresponds to $j$  reflections before reaching the static zone. The transition points $z_{j}$ are identified as
\begin{equation}
z_{j}=\frac{1}{\alpha}\left[\sqrt{(1+\alpha L_{0})^{2}+4j\alpha L_{0}}-1\right].
	\label{eq:Z_N_def}
\end{equation}
In terms of these points, we can write $n(z)$ as 
\begin{equation}
n(z)=\sum_{j=0}^{\infty}\Theta(z-z_{j}).
\label{eq:nz_asymptotic_cases}
\end{equation}
Thus, according to the diagram, the reflection number $n(z)$ is the same for any $z$ in $z_{n-1}<z\leq z_{n}$. 
By inserting the definition of $z_n$ into this relation and performing the algebraic inversion, we obtain
\begin{equation}
n-1 < \frac{(z-L_{0})(\alpha z+\alpha L_{0}+2)}{4L_{0}} \leq n.
\label{eq:inequality_n_asymptotic}
\end{equation}
By identifying this inequality with the definition of the ceiling function in Eq. (\ref{eq:ceiling_def}) and enforcing the physical condition that no reflections occur in the static region ($z \leq L_0$), we can write $n(z)$ as
\begin{equation}
	n(z)=\left\lceil \frac{(z-L_{0})(\alpha z+\alpha L_{0}+2)}{4L_{0}}\right\rceil \text{ for } z\ge L_{0},
	\label{eq:nz_asymptotic_final}
\end{equation}
and $n(z)=0$ for $z<L_0$. 
Consequently, by combining Eqs. (\ref{eq:Rz_asymptotic}) and (\ref{eq:nz_asymptotic_final}), we obtain the final exact expression for $R(z)$.
To ensure the correctness of this new analytical solution, we also subjected it to some consistency tests. First, we verified the solution indeed satisfies Moore's functional equation. Also, we compared the analytical results [Eqs. (\ref{eq:Rz_asymptotic}) and (\ref{eq:nz_asymptotic_final})] with numerical results obtained via the Dynamic Casimir package \cite{Alves-Granhen-CPC-2014}, showing perfect agreement. 
In summary, we have demonstrated how the geometric method can be applied to the case of an asymptotic contracting cavity [Eq. (\ref{eq:L-asymptotic})] to obtain a analytical solution for Moore equation.


\section{Application 3: A Resonant Motion}\label{sec:resonant_motion}

In this section, we apply the geometric method to analytically recover the first exact solution for resonant motion, originally obtained by Law~\cite{Law-PRL-1994}. The trajectory proposed by Law describes a resonant periodic motion given by Eq. (\ref{eq:Law_motion}).
Substituting this expression [Eq.~(\ref{eq:Law_motion})] into the temporal recurrence relations given by Eqs.~(\ref{t1(z)}) and
(\ref{recorrencia-ti}), and solving for the $n$-th reflection, we obtain (see Appendix \ref{apendice:recursao_temporal} for a detailed derivation)
\begin{equation}
t_{n}(z)  =  \frac{L_{0}}{2\pi}\arctan\beta_{n}+\frac{L_{0}}{2}w_{n},
\label{eq:tn_Law}
\end{equation}
where
\begin{equation}
\beta_{n}=\frac{2\tan(\kappa_{0}z)[1-(2n-1)\tan(\kappa_{0}z)\tan\theta]}{[1-(2n-1)\tan(\kappa_{0}z)\tan\theta]^{2}-\tan^{2}(\kappa_{0}z)\sec^{2}\theta},
\end{equation}
and $\kappa_{0}=\pi/L_{0}$. 
The term $w_{n}\in\mathbb{Z}$ represents the arbitrary branches arising from the inversion of the tangent function and must be adjusted to eliminate jumps, thereby ensuring the physical requirement that $R(z)$ remains a continuous function \cite{Cole-Schieve-PRA-1995,Law-PRL-1994,Lassuyt-et-al-ZMP-2025}.

\begin{figure}[t]
\epsfig{file=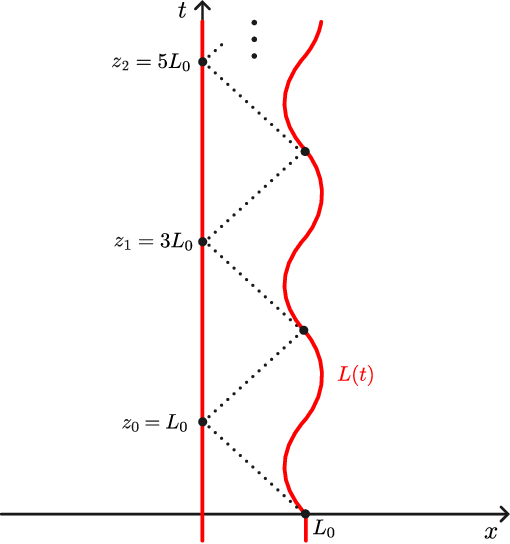,angle=00,width=1.0 \linewidth,clip=}
\caption{Evolution of the resonat cavity.
The boundary trajectories (solid red lines) represent the spacetime evolution of the cavity walls: the left boundary remains static at $x=0$, while the right follow a periodic motion. The black dotted lines represent null trajectories  that establish the boundaries of the regions responsible for counting the number of reflections $n(z)$ inside the cavity. 
}
	\label{Diagram-Law}
\end{figure}

We now turn to the determination of the discrete function $n(z)$. To this end, we depict the spacetime diagram and the corresponding null lines in Fig.~\ref{Diagram-Law}.
Following the procedure established in Sections~\ref{sec:Uniform-Motion} and~\ref{sec:gentle_contraction}, Fig.~\ref{Diagram-Law} illustrates how null trajectories (black dotted lines) partition spacetime into distinct intervals up to the static zone, with each interval corresponding to an integer value of $n(z)$.
Using the transition points between these regions, the reflection number $n(z)$ can then be written as
\begin{equation}
n(z)=\sum_{j=0}^{\infty}\Theta(z-(2j+1)L_{0}).
\label{eq:n_z_steps}
\end{equation}
Thus, according to the diagram, the reflection number $n(z)$ satisfies the following inequality
\begin{equation}
	(2n-1)L_{0} < z \leq (2n+1)L_{0},
\end{equation}
which can be rewritten as
\begin{equation}
n-1<\frac{z-L_{0}}{2L_{0}}\leq n.
\label{eq:n_isolated}
\end{equation}
Comparing Eq.~\eqref{eq:n_isolated} with the definition of the ceiling function in Eq.~\eqref{eq:ceiling_def}, we express $n(z)$ as
\begin{equation}
	n(z)=\left\lceil \frac{z-L_{0}}{2L_{0}}\right\rceil\text{ for } z \ge L_0,
	\label{eq:n_z-Law}
\end{equation}
and $n(z)=0$ for $z<L_0$. 

To eliminate discontinuities in $R(z)$ upon substituting Eqs.~(\ref{eq:n_z-Law}) and~(\ref{eq:tn_Law}) into Eq.~(\ref{R(z)-Geometric}), the branch selection in Eq.~(\ref{eq:tn_Law}) must be performed as follows (see Appendix \ref{apendice: Branches in R(z)} for a detailed derivation)
\begin{equation}
w_{n}=2\sum^{\infty}_{j=1}(-1)^{j}\Theta(z-jL_{0})+\Theta(-D)+2\Theta(D)\Theta(-N),
\label{eq:wn_definition}
\end{equation}
where $D$ and $N$ are, respectively, the denominator and numerator of $\beta_n$ in Eq. (\ref{eq:tn_Law}).

Finally, substituting Eqs.~\eqref{eq:tn_Law} and~\eqref{eq:n_z-Law} into Eq.~\eqref{R(z)-Geometric}, with $w_{n}$ given by Eq.~\eqref{eq:wn_definition}, we obtain the expression for $R(z)$ (see Appendix~\ref{apendice: Trigonometric Manipulations} for a detailed derivation).
Note that this result cannot be extended to the static region by simply setting $n(z)=0$, because the Heaviside terms at $z = L_0$, required to handle the branch cuts of the arctan function and maintain the continuity of $R(z)$, prevent the recovery of the static solution.
Therefore, the final solution becomes
\begin{eqnarray}
	R(z) &=& 2n(z)-\frac{1}{\pi}\arctan\left[\cot\frac{\pi z}{L_{0}}-\frac{2n(z)\pi\epsilon}{L_{0}}\right]+\frac{1}{2}\nonumber
	\\
	&& +\sum_{j=1}^{\infty}(-1)^{j}\Theta(z-jL_{0}) \quad\text{for } z > L_{0},
	\label{R-resonant}
\end{eqnarray}
where $n(z)$ is given by Eq. (\ref{eq:n_z-Law}) and $R(z)=z/L_{0}$ for $z\leq L_{0}$.
This expression corresponds to Law's original solution [see Eq.~\eqref{eq:Law_solution}], which was formulated in a different coordinate related to ours by $z = 2nL_0 + \xi$.
Furthermore, an important advantage of our representation is the explicit use of Heaviside terms to correctly choose the arctan branch cuts.
This mathematically formalizes the continuity condition pointed out by Law \cite{Law-PRL-1994}, whose original paper omitted these explicit terms and relied solely on a textual remark.

In conclusion, the geometric approach yields an exact analytical solution for Law's resonant motion.
Beyond providing this solution, the geometric construction also clarifies the physical meaning of the integer $n$ appearing in Law's original solution.
Law \cite{Law-PRL-1994,Law1995Tese} noted that $n$ ``somehow measures the number of round trips of light traveling inside the cavity.''
By describing the propagation of light in terms of null trajectories, we give this intuitive interpretation a precise geometrical meaning.
Specifically, we show that $n(z)$ corresponds to the exact number of reflections off the moving mirror required for the associated null trajectory to reach the static region. In general, we demonstrate that Law's solution emerges naturally from the geometry of successive null-line reflections in spacetime.


\section{Application 4: a cavity initially collapsed expanding uniformly
via a first extension of the geometric method}

\label{sec:extension-Geometric-Method-Singularity-1}

We present a first extension of the geometric method, enabling it to encompass initially collapsed cavities.
As previously discussed  in Secs.~\ref{sec:Uniform-Motion}, \ref{sec:gentle_contraction}, and \ref{sec:resonant_motion}, the geometric method proposed in 
Ref. \cite{Cole-Schieve-PRA-1995} considers an initial static region, where the function $R(z)$ is trivially known and given by Eq.~(\ref{R-static}).
Here, we present a first extension of this method, intending to describe initially collapsed cavities ($L_0=0$), which possess no prior well-defined static zone, a model investigated in Refs. \cite{Dodonov-JMP-1993,Moore-1970}. 
Using this extension, we rederive the solutions to the Moore equation, given in Eqs.~(\ref{eq:Moore-solution}) and~(\ref{eq:Dodonov/Baranov-solution}), for an initially collapsed cavity expanding uniformly.
In particular, we begin by applying the extended method to
the case where the law of motion is given by Eq. (\ref{L(t)-MRU}) with $t > -L_{0}/\alpha$ \cite{Dodonov-JMP-1993}.

\begin{figure}[H]
	\centering
	\epsfig{file=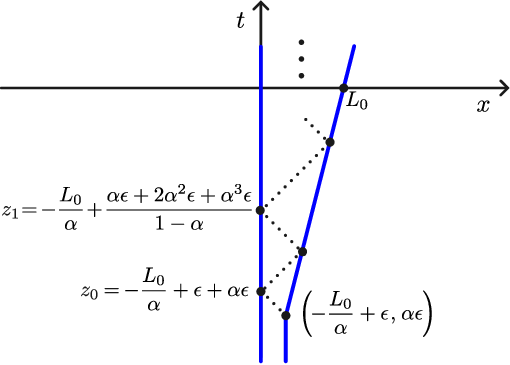, width=1.0\linewidth}
	\caption{Spacetime diagram for the uniform motion of a mirror, under the introduction of a parameter $\epsilon$ that displaces the static zone. The boundary trajectories (solid blue lines) represent the spacetime evolution of the cavity walls: the left boundary remains static at $x=0$, while the right boundary moves away at a constant velocity for $t > -L_{0}/\alpha+\epsilon$ and remains static for $t \leq -L_{0}/\alpha+\epsilon$. The black dashed lines represent null trajectories that establish the boundaries of the regions responsible for counting the number of reflections $n(z)$ inside the cavity.}
	\label{fig:Diagrama-Dodonov-Geometric}
\end{figure}

The extension is constructed by shifting the onset of the uniform motion by a small parameter $\epsilon>0$ from the singular time $t=-L_{0}/\alpha$, so that
$t_{\epsilon}=-L_{0}/\alpha+\epsilon$.
At this instant, the mirror is located at $L(t_{\epsilon})=\alpha\epsilon$.
We then assume that the mirror remains at rest at this position for $t\leq t_{\epsilon}$ and follows a uniform trajectory given by Eq.~(\ref{L(t)-MRU}) for $t>t_{\epsilon}$, as illustrated in Fig.~\ref{fig:Diagrama-Dodonov-Geometric}.
As $\epsilon\rightarrow0$, the onset of the uniform motion approaches the singularity point $(t,x)=(-L_{0}/\alpha,0)$.
Consequently, the well-defined static region disappears in this limit, recovering the initially collapsed cavity in uniform expansion, as considered by Dodonov, Klimov and Nikonov \cite{Dodonov-JMP-1993}.

We solve the cavity configuration shown in Fig.~\ref{fig:Diagrama-Dodonov-Geometric} for a finite $\epsilon>0$, in which the mirror remains at rest at $x=\alpha\epsilon$ for $t\leq -L_{0}/\alpha+\epsilon$ before entering uniform motion. The limit $\epsilon\rightarrow0$ will be taken only after obtaining the corresponding solution.
As discussed in Sec.~\ref{sec:Uniform-Motion}, the process of successive reflections ends when the null line enters the static zone (as discussed in Section \ref{Geometric-Method}).
In the static situation ($t<-L_{0}/\alpha+ \epsilon$), when the Moore equation is reduced to $R(t+\alpha\epsilon) - R(t-\alpha\epsilon) = 2$, we have the following expression for $R(z)$:  
\begin{equation}
	R(z)=z/\alpha\epsilon.
	\label{eq:R(z)-Dodonov}
\end{equation}
So, our solution to Eq. (\ref{Eq-6}) is given by 
\begin{equation}
	R(z) = 2n(z) + [t_{n} - L(t_{n})]/\alpha\epsilon.
	\label{R(z)-Geometric-Dodonov}
\end{equation}
Moreover, we know that the expression for $t_{n}(z)$ corresponding to Eq. (\ref{L(t)-MRU}) is given by Eq. (\ref{tn(z)}). Thus, substituting Eq. (\ref{tn(z)}) into  (\ref{R(z)-Geometric-Dodonov}), we have 
\begin{equation}
	R(z)=2n(z)+\frac{\left[(1-\alpha)/(1+\alpha)\right]^{n(z)}(z+L_{0}/\alpha)-L_{0}/\alpha}{\alpha\epsilon}.
	\label{eq:Rz-Dodonov}
\end{equation}
The number of reflections $n(z)$, determined according to the approach in Sec.~\ref{sec:Uniform-Motion} for the configuration in Fig.~\ref{fig:Diagrama-Dodonov-Geometric}, is given by
\begin{equation}
	n(z)=\left\lceil y(z)\right\rceil,
	\label{eq:nz_ceiling-Dodonov}
\end{equation}
where
\begin{equation}
	y(z)=\frac{\ln\left[(L_{0}+\alpha z)/(\alpha\epsilon+\alpha^{2}\epsilon)\right]}{\ln\left(1+\alpha/1-\alpha\right)}.
\end{equation}

Combining Eqs.~(\ref{eq:nz_ceiling-Dodonov}) and (\ref{eq:Rz-Dodonov}), we obtain the exact expression for $R(z)$.
We now show that, as the static zone approaches the singularity  at $t = -L_{0}/\alpha$ (that is, in the limit $\epsilon \to 0$), we can treat $n(z)$ as a continuous variable. 
Using the ceiling function property \cite{Bender1991}, given by
\begin{equation}
	\left\lceil y\right\rceil =n\Longleftrightarrow y\leq n<y+1,
	\label{eq: n_z-ceil-dodonov}
\end{equation}
and by applying this definition to our reflection number function [Eq. (\ref{eq:nz_ceiling-Dodonov})], we obtain the inequality
\begin{equation}
	y(z)\leq n(z)<y(z)+1.
\end{equation}
Dividing all terms of the inequality by the  term on the left yields
\begin{equation}
	1\leq n(z)/y(z)<1+1/y(z).
	\label{eq:n(z)-desiqualdade}
\end{equation}
Given that 
\begin{equation}
	\lim_{\epsilon\to0}\frac{1}{y(z)}=0.
\end{equation}
Therefore, it follows from the squeeze theorem applied to Eq. (\ref{eq:n(z)-desiqualdade}) that the ceiling function can be omitted as 
\begin{equation}
	n(z)\approx\frac{\ln\left[(L_{0}+\alpha z)/(\alpha\epsilon+\alpha^{2}\epsilon)\right]}{\ln\left[(1-\alpha)/(1+\alpha)\right]},
	\label{eq:n(z)-cont-Dodonov}
\end{equation}
that is, the number of reflections $n(z)$ can be treated as a continuous variable.
Substituting Eq. (\ref{eq:n(z)-cont-Dodonov}) into (\ref{eq:Rz-Dodonov}), we obtain 
\begin{equation}
	R(z)=2\frac{\ln\left[(L_{0}+\alpha z)/(\alpha\epsilon+\alpha^{2}\epsilon)\right]}{\ln\left[(1-\alpha)/(1+\alpha)\right]}+\frac{1+\alpha}{\alpha}-\frac{L_{0}}{\alpha^{2}\epsilon}.
	\label{eq:Rz-sub-n(z)}
\end{equation}
As we let $\epsilon \to 0$ in Eq.~(\ref{eq:Rz-sub-n(z)}) (to recover the uniform expansion model of Dodonov, Klimov and Nikonov \cite{Dodonov-JMP-1993} for $t> -L_{0}/\alpha$), 
we note the appearance of divergent terms.
To address this issue, we perform a phase redefinition in the field modes, renormalizing the solution to obtain a physically consistent expression \cite{Baranov-Shirokov-SPJ-1968}.
To do this, we decompose $R(z)$ into a finite part and a divergent part:
\begin{equation}
	\mathcal{R}(z)=R_{0}(z)+R_{\epsilon},
	\label{eq:R(z)-divid}
\end{equation}
where
\begin{equation}
	R_{0}(z)=2\frac{\ln\left(1+\alpha z/L_{0}\right)}{\ln\left[(1-\alpha)/(1+\alpha)\right]}+\frac{1}{\alpha}+1,
	\label{eq:R(z)-alpha-Div}
\end{equation}
and
\begin{equation}
	R_{\epsilon}=2\frac{\ln\left[(\alpha\epsilon+\alpha^{2}\epsilon)/L_{0}\right]}{\ln\left[(1-\alpha)/(1+\alpha)\right]}-\frac{L_{0}}{\alpha^{2}\epsilon}.
\end{equation}
Substituting Eq.~(\ref{eq:R(z)-divid}) into the field solutions (\ref{eq:field-1}) and (\ref{eq:field-2}), we absorb the divergent phase factor into the field normalization through the redefinition of the field operators, as follows:
\begin{equation}
	\widetilde{a}_{n}=a_{n}e^{-in\pi R_{\epsilon}}.
\end{equation}

Consequently, the solution to the field becomes 
\begin{equation}
	\phi(t,x)=\sum_{n=1}^{\infty}\tilde{a}_{n}\tilde{\phi}_{n}(t,x)+\tilde{a}_{n}^{\dagger}\tilde{\phi}_{n}^{*}(t,x),
\end{equation}
where
\begin{equation}
	\tilde{\phi}(t,x)=\frac{i}{2\sqrt{n\pi}}\left[e^{-in\pi R_{0}(v)}-e^{-in\pi R_{0}(u)}\right].
\end{equation}

Hence, we eliminate the singularity that arises in the limit $\epsilon \to 0$, and the solution to Moore equation is provided by Eq. (\ref{eq:R(z)-alpha-Div}).

As in the previous case, a phase redefinition of the field modes is required to remove the divergence in Eq. (\ref{eq:R(z)-alpha-Div}) as $\alpha \to 0$.
This new phase renormalization ensures that, when taking the static limit, we recover the usual solution for mirrors at rest [Eq. (\ref{R-static})] \cite{Baranov-Shirokov-SPJ-1968}.
Additionally, we include the constant term in the renormalization of Eq. (\ref{eq:R(z)-alpha-Div}), since Moore equation is satisfied up to an additive constant. As a result, the final solution for $R(z)$ assumes the form presented in Eq.~(\ref{eq:Dodonov/Baranov-solution}).

We now consider the second case investigated by Moore \cite{Moore-1970},
where the law of motion is given by $L(t)=\alpha t$ for $t > 0$, as illustrated in Fig. \ref{fig:Fig-literature}a. 
To recover this model from Eq.~(\ref{L(t)-MRU}), we introduce a time shift $(t'=t+L_0/\alpha)$ in Eq.~(\ref{L(t)-MRU}), under which the uniform trajectory become $L(t^{\prime})=\alpha t^{\prime}$.
Consequently, the spacetime diagram in Fig. \ref{fig:Diagrama-Dodonov-Geometric} reflects this shift, positioning the static mirror at $x=\alpha\epsilon$ for $t^{\prime} \leq \epsilon$.
By proceeding with the calculations in a manner analogous to the previous case, the solution expressed in Eq. (\ref{eq:Moore-solution}) is reobtained.
Thus, we have shown that the geometric method can be extended to initially collapsed cavities, which have no prior well-defined static zone, by introducing an auxiliary static region controlled by the parameter $\epsilon$ and subsequently taking the limit  $\epsilon\rightarrow0$. 
In this limit, the static region is suppressed and the original collapsed-cavity configuration is recovered.
With this extension, we recover the analytical solutions for initially collapsed cavities expanding uniformly, encompassing both Moore's model \cite{Moore-1970} (Fig. \ref{fig:Fig-literature}a) and the model investigated by Dodonov, Klimov, and Nikonov, where the initial collapse occurs at $t = -L_0/\alpha$ \cite{Dodonov-JMP-1993}.
Therefore, this extension broadens the applicability of the geometric method, paving the way for the exploration of new models of initially collapsed cavities.

Furthermore, this discussion also shows that we can always construct solutions for Moore equation by shifting our static zone to other points in spacetime that allow for calculating the number of reflections. This observation applies equally to the other models we present throughout the paper.

\section{Applications 5: Combined laws of motion via a Second Extension of the Geometric Method}
\label{sec: extension-Geometric-Method-2}

\begin{figure}[H]
	\centering
	\epsfig{file=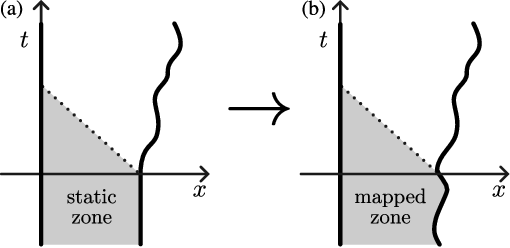, width=1.0\linewidth}
	\caption{(a) Representation of the static zone (shaded area), where $R(z)$ is given by Eq.~(\ref{R-static}) and used by Cole and Schieve \cite{Cole-Schieve-PRA-1995} in Eq.~\eqref{R(z)-Geometric}. (b) Representation of the mapped zone (shaded area), where $R(z)$ is assumed to be known, although not given by Eq.~(\ref{R-static}).
	}
	\label{fig:Fig-no-static-zone}
\end{figure}

The geometric method developed by Cole and Schieve \cite{Cole-Schieve-PRA-1995} was originally formulated under the requirement of an initial static region [see Fig. \ref{fig:Fig-no-static-zone}a], where the field was not affected by the mirror motion and the corresponding solution to the Moore equation is given by Eq.~(\ref{R-static}). 
Here, we present another extension of this method, demonstrating that new exact analytical solutions to the Moore equation can be constructed 
from any spacetime region (hereafter termed mapped zone) where  $R(z)$ has already been determined, even by other methods [see Fig. \ref{fig:Fig-no-static-zone}b].

To formulate this extension, let us consider that initially the mirror follows a generic law of motion for which $R(z)$ is already known [Fig. \ref{fig:Fig-no-static-zone}b], namely $R(z)=R_{\text{map}}(z)$.
At a certain instant, the mirror adopts a new trajectory $L(t)$, for which the corresponding function $R(z)$ becomes undetermined.
Following the construction introduced in Sec.~\ref{Geometric-Method}, we analytically determine the solution for the new trajectory by tracing a sequence of null lines that undergo successive reflections at the cavity boundaries.
However, unlike the original geometric method, the backward tracing terminates
as soon as the null line enters the spacetime region where the solution to the Moore equation is already known and given by $R_{\text{map}}(z)$, that is, in the mapped zone.
In this way, while the original Cole-Schieve construction \cite{Cole-Schieve-PRA-1995} replaces Eq.~\eqref{Eq-6} by Eq.~\eqref{R(z)-Geometric}, our extension replaces it by
\begin{equation}
	R(z)=2n(z)+R_{\text{map}}\left(t_{n}(z)-L(t_{n})\right),
	\label{eq:R(z)-Knowledge}
\end{equation}
where $n(z)$ denotes the number of reflections off the moving mirror required to connect the null line $v=z$ to a null line in the mapped zone, and the function $t_{n}(z)$ is still given by Eqs.~(\ref{t1(z)}) and~(\ref{recorrencia-ti}).
In this way, Eq.~(\ref{eq:R(z)-Knowledge}) generalizes Eq.~(\ref{R(z)-Geometric}), allowing to obtain
$R(z)$ for combinations of different laws of motion. 
Next, in Secs.~\ref{subsec:Uniform Cavity Expansion with an Abrupt Halt}, \ref{subsec:Uniform Cavity Expansion with a Subsequent Resonant Motion}, \ref{subsec:Resonant Cavity Semicycle}, and \ref{subsec:Modulated Resonant Oscillation}, we illustrate  four applications of
this extension.
 
%
\subsection{Application 5.1: Uniform Cavity Expansion with an Abrupt Halt}
\label{subsec:Uniform Cavity Expansion with an Abrupt Halt}
\begin{figure}[!ht]
	\centering
	\epsfig{file=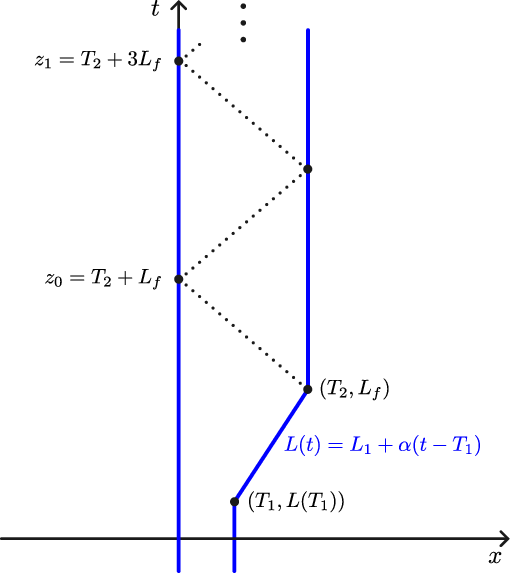, width=1.0\linewidth}
	\caption{Evolution of the uniform cavity expansion with an abrupt halt. The solid blue lines represent the trajectories of the cavity boundaries: the left mirror remains static at $x=0$, while the right mirror is initially at rest, executes a uniform motion, and abruptly returns to rest. The dotted lines represent null trajectories establishing the boundaries of the regions used for counting the number of reflections, $n(z)$ inside the cavity until reaching the mapped zone.}
	\label{fig:Diagrama-MRU-STOP}
\end{figure}

As an application of this extension, we will construct a new exact analytical solution to the Moore equation in a scenario in which the mirror starts from rest, moves uniformly for a finite time interval, and then abruptly returns to rest, as illustrated in Fig.~\ref{fig:Diagrama-MRU-STOP}. This law of motion is described by
\begin{equation}
	L(t)=\begin{cases}
		L_{1}, & t<T_{1},\\
		L_{1}+\alpha(t-T_{1}), & T_{1}\leq t<T_{2},\\
		L_{f}, & t\geq T_{2},
	\end{cases}
	\label{eq: MRU-STOP}
\end{equation}
where $L_{f}=L_{1}+\alpha(T_{2}-T_{1})$.
To construct this solution, we first consider the  configuration in which the mirror follows the same initial static and uniform-motion stages but does not undergo the final stop:
\begin{equation}
	L(t)=\begin{cases}
		L_{1}, & t<T_{1}, \\
		L_{1}+\alpha(t-T_{1}), & T_{1}\leq t.
	\end{cases}\label{eq: MRU-1}
\end{equation}
The solution to the Moore equation for this  motion can be obtained following the same procedure described in Sec.~\ref{sec:Uniform-Motion}, the only difference being that the initial static cavity now is given by $\ensuremath{R_{\text{static}}(z)=z/L_{1}}$. Thus, we obtain
\begin{equation}
	R(z)=2n_{2}(z)+\left(\frac{1-\alpha}{1+\alpha}\right)^{n_{2}(z)}\left[\frac{(z-T_{1})}{L_{1}}+\frac{1}{\alpha}\right]+\frac{T_{1}}{L_{1}}-\frac{1}{\alpha},\label{eq: R(z)-MRU-1}
\end{equation}
where
\begin{equation}
	n_{2}(z)=\begin{cases}
		0 & z\leq T_{1}+L_{1}\\
		\left\lceil \frac{\ln\left[\frac{\alpha(z-T_{1})+L_{1}}{L_{1}(1+\alpha)}\right]}{\ln\left[(1+\alpha)/(1-\alpha)\right]}\right\rceil  & z>T_{1}+L_{1}
	\end{cases}.
	\label{eq: n(z)-MRU-1}
\end{equation}

We now use this analytical solution to construct the solution after the abrupt stop at $(T_{2}, L_{f})$.
Specifically, the preceding dynamical regime is taken as the \textit{mapped zone}, as illustrated in Fig.~\ref{fig:Diagrama-MRU-STOP}.
In this region, the function $R_{\text{map}}(z)$ is therefore given by Eqs.~(\ref{eq: R(z)-MRU-1}) and~(\ref{eq: n(z)-MRU-1}).

From the geometric construction shown in Fig.~\ref{fig:Diagrama-MRU-STOP}, we determine the number of reflections $n_{1}(z)$ required to reach this mapped zone, together with the argument $\gamma_{1}(z) = t_{n} - L(t_{n})$, following a procedure analogous to those described in Secs.~\ref{sec:Uniform-Motion}, \ref{sec:gentle_contraction}, and \ref{sec:resonant_motion}.
Finally, substituting these results into Eq.~(\ref{eq:R(z)-Knowledge}), with $R_{\text{map}}(z)$  given by Eqs.~(\ref{eq: R(z)-MRU-1}) and~(\ref{eq: n(z)-MRU-1}), our final solution becomes
\begin{equation}
	\begin{split}R(z)={} & 2n_{1}(z)+2n_{2}(\gamma_{1}(z))+\frac{T_{1}}{L_{1}}-\frac{1}{\alpha}\\
		& +\left(\frac{1-\alpha}{1+\alpha}\right)^{n_{2}\left(\gamma_{1}(z)\right)}\left[\frac{\left(\gamma_{1}(z)-T_{1}\right)}{L_{1}}+\frac{1}{\alpha}\right],
	\end{split}
	\label{eq: R(z)-Final-Knowlegde}
\end{equation}
where
\begin{equation}
	\gamma_{1}(z)=z-2n_{1}(z)L_{f},
	\label{eq: gamma(z)}
\end{equation}
\begin{equation}
	n_{1}(z)=\begin{cases}
		0, & z\leq T_{2}+L_{f},\\
		\left\lceil \frac{z-T_{2}-L_{f}}{2L_{f}}\right\rceil , & z>T_{2}+L_{f},
	\end{cases}\label{eq: n1-final}
\end{equation}
and the function $n_{2}(z)$ was defined in Eq. (\ref{eq: n(z)-MRU-1}).
The Eqs.~(\ref{eq: R(z)-Final-Knowlegde}) through~(\ref{eq: n1-final}) satisfy the Moore equation along the entire mirror trajectory described Eq.~(\ref{eq: MRU-STOP}) and provide a general solution for a mirror that is initially at rest, undergoes uniform motion, and then comes to an abrupt stop.
For the particular case in which the system variables are parameterized by three independent quantities, $\tau$, $\alpha_{1}$, and $\alpha_{2}$, with  $L_{1}=\tau/\alpha_{1}$, $T_{1}=\tau\left(1-1/\alpha_{1}\right)$, $T_{2}=\tau\left(1+1/\alpha_{2}\right)$, and $\alpha=(\alpha_{1}-\alpha_{2})/(\alpha_{1}+\alpha_{2})$,
we recover Moore's law of motion \cite{Moore-1970} presented in Eq.~(\ref{eq:Moore-lei-MRU}). 
Under this parametrization, the general solution given
by Eqs.~(\ref{eq: R(z)-Final-Knowlegde}) to~(\ref{eq: n1-final}) simplifies to
\begin{equation}
	R(z)=\begin{cases}
		\alpha_{1}(z-\tau)/\tau+\alpha_{1}, & z\leq\tau,\\
		\alpha_{2}(z-\tau)/\tau+\alpha_{1}, & z>\tau.
	\end{cases}\label{eq: R(z)-Moore-Reobtain}
\end{equation}
Thus, we recover Moore's exact analytical solution given by Eq.~(\ref{eq:R(z)-Moore-MRU}), up to the additive constant $\alpha_{1}$.
It is worth noting that every solution to the Moore equation is satisfied to within a constant factor.
\subsection{Application 5.2: Uniform Cavity Expansion with a Subsequent Resonant Motion}
\label{subsec:Uniform Cavity Expansion with a Subsequent Resonant Motion}

In addition to the scenario of uniform cavity expansion with an abrupt halt, the extension of the geometric method allows for the construction of new exact solutions to the Moore equation by coupling the uniform expansion regime to different subsequent dynamics.
This construction is possible provided that the new dynamical regime shares the same \textit{mapped zone}, where the solution is given by Eqs.~(\ref{eq: R(z)-MRU-1}) and~(\ref{eq: n(z)-MRU-1}), and that the law of motion adopted after the transition allows the temporal recurrence relations [Eqs.~(\ref{t1(z)}) and~(\ref{recorrencia-ti})] to be inverted analytically.

\begin{figure}[!ht]
	\centering
	\epsfig{file=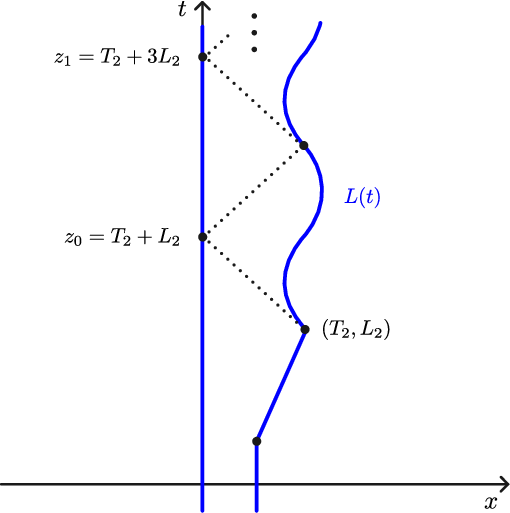, width=1.0\linewidth}
	\caption{Evolution of the uniform cavity expansion with a subsequent resonant motion. The solid blue lines represent the trajectories of the cavity boundaries: the left mirror remains static at $x = 0$, while the right mirror is initially at rest, executes a uniform motion, and subsequently transitions into a resonant dynamical regime . The dotted lines represent null trajectories establishing the boundaries of the regions used for counting the number of reflections $n(z)$ inside the cavity until reaching the mapped zone.}
	\label{fig:Diagrama-MRU-LAW}
\end{figure}

As an example, we consider the scenario in which the mirror, executing uniform expansion, abruptly transitions to Law's resonant oscillatory motion \cite{Law-PRL-1994}, as illustrated in Fig.~\ref{fig:Diagrama-MRU-LAW}.The corresponding law of motion is

\begin{equation}
	L(t)=\begin{cases}
		L_{1}, & t<T_{1}\\
		L_{1}+\alpha(t-T_{1}), & T_{1}\leq t<T_{2}\\
		L_{\text{osc}}(t), & t\geq T_{2}
	\end{cases},
	\label{eq: L(t)-Knowledege-2}
\end{equation}
where the trajectory in the resonant regime is governed by
\begin{equation}
	\begin{split}
		L_{\text{osc}}(t) ={} & L_{2}\bigg\{ 1-\frac{\theta}{2\pi} \\
		& + \frac{1}{2\pi}\arcsin\left(\sin\theta\cos\left[\frac{2\pi(t-T_{2})}{L_{\text{2}}}\right]\right)\bigg\},
	\end{split}
	\label{eq: L-osc-split}
\end{equation}
with $L_{2}=L_{1}+\alpha\left(T_{2}-T_{1}\right)$\;\text{and}\;$\theta=\arctan(\epsilon\pi/L_{2})$.
Using the geometric construction shown in Fig.~\ref{fig:Diagrama-MRU-LAW} and following the same procedure described in Sec.~\ref{sec:resonant_motion}, we determine the number of reflections $n_{3}(z)$ required for a null line to reach the mapped zone, together with the argument $\gamma_{2}(z)=t_{n}-L(t_n)$, obtaining
\begin{equation}
	n_{3}(z)=\begin{cases}
		0, & z\leq T_{2}+L_{2},\\
		\left\lceil \frac{z-T_{2}-L_{2}}{2L_{2}}\right\rceil , & z>T_{2}+L_{2},
	\end{cases}\label{eq:n_3(z)}
\end{equation}
and
\begin{equation}
	\begin{split}\gamma_{2}(z)={} & T_{2}-\frac{L_{2}}{\pi}\arctan\left[\cot\left(\frac{\pi(z-T_{2})}{L_{2}}\right)-\frac{2n_{3}(z)\pi\epsilon}{L_{2}}\right]\\
		& +L_{2}\left(\left\lfloor \frac{z-T_{2}}{L_{2}}\right\rfloor -2n_{3}(z)+\frac{1}{2}\right).
	\end{split}
	\label{eq: gamma-2}
\end{equation}
Finally, by substituting Eqs.~(\ref{eq:n_3(z)}) and~(\ref{eq: gamma-2}), together with the solution \(R_{\text{map}}(z)\), previously obtained from Eqs.~(\ref{eq: R(z)-MRU-1}) and~(\ref{eq: n(z)-MRU-1}), into the general expression given by Eq.~(\ref{eq:R(z)-Knowledge}), 
we obtain a new exact analytical solution that satisfies to the Moore equation across the three motion regimes described in Eq.~(\ref{eq: L(t)-Knowledege-2}).
This result also illustrates the importance of recovering exact solutions known in the literature through the geometric method, such as Law's resonant solution \cite{Law-PRL-1994} derived in Sec.~\ref{sec:resonant_motion},
since the corresponding analytical expressions for $n(z)$ and $\gamma(z)$ are  what makes this coupling analytically possible.
%

\subsection{Application 5.3: Oscillation Cavity Semicycle with an Abrupt Halt}
\label{subsec:Resonant Cavity Semicycle}

Another key feature of the extended geometric method is its ability to construct exact analytical solutions in configurations where null lines, traced backward from a region in which $R(z)$ is unknown, may intersect multiple mapped zones.

\begin{figure}[!ht]
	\centering
	\epsfig{file=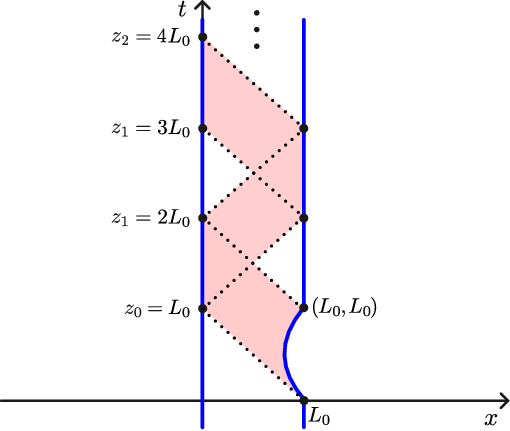, width=1.0\linewidth}
	\caption{Evolution of the oscillation cavity semicycle with an abrupt halt. The solid blue lines represent the trajectories of the cavity boundaries: the left mirror remains static at $x = 0$, while the right mirror is initially at rest, executes a half-cycle of oscillation and then comes to an abrupt halt. The dotted lines represent null trajectories that define the boundaries of the regions used to count the number of reflections $n(z)$ inside the cavity until they reach one of the mapping zones. The shaded areas highlight the domains of spacetime in which the null lines are mapped to the dynamic zone $L_{0} < z \leq 2L_{0}$.}
	\label{fig:Oscillation Cavity Semicycle with an Abrupt Halt}
\end{figure}

As an example, in this subsection we consider a dynamical model in which an initially static mirror undergoes a half-cycle of oscillation and then comes to an abrupt halt, as illustrated in Fig.~\ref{fig:Oscillation Cavity Semicycle with an Abrupt Halt}. 
This law of motion is described by
\begin{equation}
	L(t)=\begin{cases}
		L_{0}, & t<0,\\
		L_{\text{dyn}}(t), & 0\leq t<L_{0},\\
		L_{0}, & t\geq L_{0},
	\end{cases}\label{eq: Half-Oscillation-Stop}
\end{equation}
where $L_{\text{dyn}}(t)$ is given by Eq. (\ref{eq:Law_motion}).
As illustrated in Fig.~\ref{fig:Oscillation Cavity Semicycle with an Abrupt Halt}, tracing the null trajectories backward until they reach the region $z\leq2L_{0}$ reveals two distinct mapped regions.
This distinction reflects the structure of Law's solution for $R(z)$, which is composed of a static and a dynamical part, as discussed in Sec.~\ref{sec:resonant_motion}.
For $z \leq L_{0}$, the null trajectories reach the static region, where $R_{\text{map}}(z)$ is given by Eq.~(\ref{R-static}).
In contrast, for $L_{0} < z \leq 2L_{0}$, the null trajectories reach a region that has already been affected by the mirror oscillation, where $R_{\text{map}}(z)$ is given by Eq.~(\ref{R-resonant}) with $n(z)=1$.
As show in Fig;~\ref{fig:Oscillation Cavity Semicycle with an Abrupt Halt} null trajectories traced from $(2m+1)L_0 < z \leq (2m+2)L_0$ (with $m=1,2,\dots$), terminate in the mapped region $L_{0} < z \leq 2L_{0}$, whereas those originating from $2mL_0 < z \leq (2m+1)L_0$, reach the static mapped region $z \leq L_{0}$.
Following the same geometric construction used in Secs.~\ref{sec:Uniform-Motion}, \ref{sec:gentle_contraction}, and~\ref{sec:resonant_motion}, we determine the number of reflections $n_{4}(z)$ required to enter the region $z \leq 2L_0$ and map one of the two solutions of $R_{\text{map}}$ and the corresponding argument $\gamma_{3}(z)=t_{n}-L(t_n)$. We obtain the global solution

\begin{subequations}
	\begin{equation}
		R(z)=\begin{cases}
			R_{\text{static}}(z), & z\leq L_{0},\\
			R_{\text{dyn}}(z), & (2m-1)L_{0}<z\leq2mL_{0},\\
			R_{\text{static}}(z), & 2mL_{0}<z\leq(2m+1)L_{0},
		\end{cases}\label{eq: R(z)-Dual}
	\end{equation}
	where $m=1,2,\dots$, $R_{\text{static}}$ is given by Eq.~(\ref{R-static}) and
	\begin{equation}
		\begin{split}R_{\text{dyn}}(z) & =2n_{4}(z)+\frac{5}{2}-\frac{1}{\pi}\arctan\left(\cot\frac{\pi z}{L_{0}}-\frac{2\pi\epsilon}{L_{0}}\right)\\
			& \quad+\sum_{j=1}^{2}(-1)^{j}\Theta\big(z-2n_{4}(z)L_{0}-jL_{0}\big),
		\end{split}
		\label{eq: R-dyn}
	\end{equation}
	with
	\begin{equation}
		n_{4}(z)=\begin{cases}
			0, & z\leq2L_{0},\\
			\left\lceil \frac{z-2L_{0}}{2L_{0}}\right\rceil , & z>2L_{0}.
		\end{cases}\label{eq:n4(z)}
	\end{equation}
	\label{eq:R-Full-System}
\end{subequations}
The Eq.~(\ref{eq:R-Full-System}) satisfies the Moore equation throughout all three dynamical regimes defined by Eq.~(\ref{eq: Half-Oscillation-Stop}).
Furthermore, we compared the analytical results [Eq. (\ref{eq:R-Full-System})] with the numerical results obtained using the Dynamic Casimir package \cite{Alves-Granhen-CPC-2014}, and found perfect agreement. 
This result also geometrically demonstrates that, in the white regions shown in Fig.~\ref{fig:Oscillation Cavity Semicycle with an Abrupt Halt}, where the field is not affected by the mirror's oscillation, the solution remains governed solely by its static form [Eq.~\ref{R-static}].
In contrast, in the highlighted (shaded) regions, the dynamics of the field carry information about the mirror's oscillating trajectory, always including the contribution of the analytical term $R_{\text{dyn}}$.

Therefore, this example demonstrates how to construct exact analytical solutions to the Moore equation by connecting more than one mapped zone, $R_{\text{map}}(z)$. 
More broadly, this result shows that the extension of the geometric method can be applied to even more complex scenarios, allowing us to solve dynamics in domains where the function $R_{\text{map}}$ is only partially known or to model systems with more distinct mapped regions.

\subsection{Application 5.4: Modulated Cavity Resonant Oscillation with a Subsequent Asymptotic Contraction}
\label{subsec:Modulated Resonant Oscillation}
A further advantage of this extension of the geometric method is that the function $R_{\text{map}}(z)$ adopted in the initial mapped zone is independent of the technique used to derive it.
Therefore, laws of motion with corresponding exact analytical solutions to the Moore equation obtained by other methods in the literature \cite{Moore-1970, Dodonov-JMP-1993, Law-PRL-1994, Dodonov1996Book,Law1995Tese,Baranov-Shirokov-SPJ-1968, Fulling-Davies-PRSA-1976, Castagnino-Ferraro-AnnPhys-1984, Wu-PRA-1999, Li-Ping-Fu-PRA-1997, Lassuyt-et-al-ZMP-2025}, such as those reviewed in Sec.~\ref{sec:Literature-Overview}, can be adopted to describe the system's initial dynamics, with the associated solutions identified as $R_{\text{map}}(z)$.
Consequently, subsequent laws of motion can be coupled to these prior regimes, provided that the new trajectories allow an analytical construction through the geometric method.

\begin{figure}[t]
	\centering
	\epsfig{file=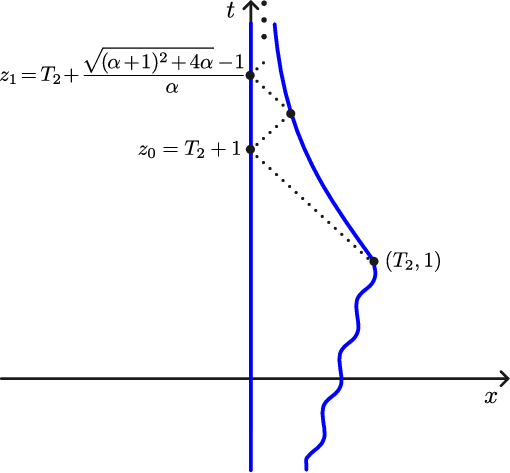, width=1.0\linewidth}
	\caption{Evolution of the modulated cavity resonant oscillation with a subsequent asymptotic contraction. The solid blue lines represent the trajectories of the cavity boundaries: the left mirror remains static at $x = 0$, while the right mirror is initially at rest, executes a modulated resonant motion, and subsequently transitions into an asymptotic contraction regime for $t \geq T_2$. The dotted lines represent null trajectories establishing the boundaries of the regions used for counting the number of reflections $n(z)$ inside the cavity until reaching the mapped zone.}
	\label{fig:Modulated Resonant with a Subsequent Asymptotic Contraction}
\end{figure}

As an example, we consider the modulated resonant oscillatory law of motion developed by Fu, Duan, and Guo \cite{Li-Ping-Fu-PRA-1997} to describe the initial dynamics of the cavity, as presented in Eq.~(\ref{eq:L(t)-Fu}).
As a result, its corresponding solution to the Moore equation Eq.~(\ref{eq:R(z)-Fu}), is adopted as $R_{\text{map}}(z)$.
By choosing to couple this regime to the asymptotic contraction motion discussed in Sec.~\ref{sec:gentle_contraction}, we obtain the global law of motion illustrated in Fig.~\ref{fig:Modulated Resonant with a Subsequent Asymptotic Contraction}, given by
\begin{equation}
	L(t)=\begin{cases}
		\dfrac{1}{1+\lambda}, & t<-\dfrac{1}{1+\lambda},\\[3ex]
		\dfrac{1+\lambda t'(t)}{1+\lambda}+Q(t'(t)), & -\dfrac{1}{1+\lambda}\le t\le1,\\[3ex]
		1, & 1<t<T_{2},\\[3ex]
		\dfrac{1}{1+\alpha(t-T_{2})}, & t\ge T_{2},
	\end{cases}
	\label{eq: L(t)-coupled_modular_contraction}
\end{equation}
where $t'(t)=[t(1+\lambda)+1]/(2+\lambda)$
and $
\ensuremath{Q(t^{\prime})=-\left[\lambda/m\pi(2+\lambda)\right]\sin\left[2m\pi t^{\prime}+m\pi Q(t^{\prime})\right].}$
It is crucial to note that the coupled contraction motion only begins at a later time, $T_{2} > 1$. 
Consequently, any null line traced from the region $z > T_{2} + 1$ will intersect the mapped zone at $z > 0$.
For this reason, we solely consider the solution to Eq.~(\ref{eq:R(z)-Fu}) in the domain $z > 0$ to represent $R_{\text{map}}(z)$.

Using the geometric construction shown in Fig.~\ref{fig:Modulated Resonant with a Subsequent Asymptotic Contraction}, and
following the same procedure described in Secs.~\ref{sec:Uniform-Motion},\ref{sec:gentle_contraction}, and~\ref{sec:resonant_motion}, we determine the number of reflections $n_{5}(z)$ required to enter the mapped zone, together with the argument $\gamma_{4}(z)=t_{n}-L(t_n)$.
By substituting these results into the general expression of Eq.~(\ref{eq:R(z)-Knowledge}),
with $R_{\text{map}}(z)$ specified by Eq.~(\ref{eq:R(z)-Fu}), we obtain the exact analytical solution
\begin{equation}
	R(z)=\begin{cases}
		(1+\lambda)z, & z\le0,\\[1.5ex]
		2n_{5}(z)+\gamma_{4}(z)+\dfrac{\lambda}{m\pi}\sin(m\pi\gamma_{4}(z)), & z>0,
	\end{cases}
	\label{eq:R(z)-Fu-Geometric}
\end{equation}
where
\begin{equation}
	n_{5}(z)=\begin{cases}
		0, & z\leq T_{2}+1,\\
		\left\lceil \frac{[\alpha(z-T_{2})+1]^{2}-(\alpha+1)^{2}}{4\alpha}\right\rceil , & z>T_{2}+1,
	\end{cases}\label{eq: n_=00007B4=00007D(z)}
\end{equation}
and
\begin{equation}
	\gamma_{4}(z)=T_{2}+\frac{-1+\sqrt{[\alpha(z-T_{2})+1]^{2}-4\alpha n_{4}(z)}}{\alpha}.
	\label{eq:gamma_{3}(z)}
\end{equation}

The analytical solution above satisfies to the Moor equation along the entire trajectory described in Eq.~(\ref{eq: L(t)-coupled_modular_contraction}).

In this way, we have shown how to construct new exact solutions to the Moore equation by coupling trajectories to initial dynamic regimes whose corresponding functions $R_{\text{map}}(z)$ were not necessarily derived using the geometric method.
Therefore, the results established in this section demonstrate that the extension of the geometric method, synthesized in Eq.~(\ref{eq:R(z)-Knowledge}), acts as a generator of infinite exact analytical solutions capable of coupling multiple regimes of motion.

For instance, it suffices to assume that the final solutions for $R(z)$ obtained in Secs.~\ref{subsec:Uniform Cavity Expansion with an Abrupt Halt}, \ref{subsec:Uniform Cavity Expansion with a Subsequent Resonant Motion}, \ref{subsec:Resonant Cavity Semicycle} and~\ref{subsec:Modulated Resonant Oscillation}, act as the new function $R_{\text{map}}(z)$.
Starting from an arbitrary transition instant, a new law of motion $L(t)$ is coupled, and the standard geometric procedures are applied to determine the solution to the Moore equation in this new regime.
This procedure can be repeated iteratively, with each newly obtained solution becoming the \(R_{\text{map}}(z)\) for the next dynamical regime.

Thus, the extended geometric method enables the systematic construction of exact analytical solutions involving an infinite number of coupled laws of motion, provided that each subsequent dynamics remains analytically tractable within the geometric construction.
To illustrate a diversity of alternative configurations attainable through our extension, Fig.~\ref{fig:4-intro-extension}
presents representative examples from the infinite class of exact analytical solutions that can, in principle, be constructed.
%
%


\section{Final Comments}
\label{final-comments}

In this paper, we have demonstrated that the geometric method proposed by Cole and Schieve \cite{Cole-Schieve-PRA-1995}, which has 
traditionally been regarded solely as a numerical method, is also a powerful tool for deriving exact analytical solutions to the Moore equation.
By applying this geometric method, we have obtained novel exact solutions for two classes of laws of motion 
starting from rest, with well-defined static zone: an expanding cavity with a uniformly moving mirror [Figs.~\ref{fig:Nossos-movimentos}(a) and \ref{Linhas-nulas-MRU}; Eqs.~(\ref{eq:Rz_inicial}) and (\ref{n(z)-ceil})], and a contracting cavity with a mirror that approaches the static one asymptotically [Figs.~\ref{fig:Nossos-movimentos}b and \ref{Diagrama-Suave}; Eqs.~(\ref{eq:Rz_asymptotic}) and (\ref{eq:nz_asymptotic_final})].
In addition, we recovered the first exact analytical solution to the Moore equation for a class of resonant mirror motions  [Figs.~\ref{fig:Nossos-movimentos}c and \ref{Diagram-Law}; Eqs.~(\ref{R-resonant}) and (\ref{eq:n_z-Law})], originally obtained by Law \cite{Law-PRL-1994}.
With these applications, we showed that when the temporal recurrence relations [Eqs.~(\ref{t1(z)}) and (\ref{recorrencia-ti})] have analytical exact solution, the function $R(z)$ admits an exact analytical solution. 

Furthermore, we presented two extensions to the original geometric method proposed by Cole and Schieve \cite{Cole-Schieve-PRA-1995}.
First, we enabled the method to describe initially collapsed cavities, which possess no prior well-defined static zone.
By doing so, we recovered the analytical solutions for classes of laws of motion describing initially collapsed mirrors in uniform motion [Fig.~\ref{fig:Fig-literature}a], encompassing both Moore's model \cite{Moore-1970} [Eq.~(\ref{eq:Moore-lei-MRU})] and that investigated by Dodonov, Klimov and Nikonov \cite{Dodonov-JMP-1993}, in which the cavity is collapsed for $t \leq -L_0/\alpha$ [Eq.~(\ref{eq:R(z)-alpha-Div})].
Se\-cond, we demonstrated that new exact solutions can be constructed from any spacetime region where the function $R(z)$ 
has been previously determined, even by other methods (see Fig.~\ref{fig:Fig-no-static-zone}).
By applying this extension, we obtained exact analytical solutions for four novel classes of mirror trajectories involving combinations of different motion regimes: a uniform expansion followed by an abrupt halt [Figs.~\ref{fig:3-intro-extension}b and \ref{fig:Diagrama-MRU-STOP}; Eqs.~(\ref{eq: R(z)-Final-Knowlegde})--(\ref{eq: n1-final})], a uniform expansion followed by resonant motion
[Figs.~\ref{fig:3-intro-extension}c and \ref{fig:Diagrama-MRU-LAW}],
a single oscillation semicycle followed by an abrupt halt [Figs.~\ref{fig:3-intro-extension}d and \ref{fig:Oscillation Cavity Semicycle with an Abrupt Halt}; Eq.~(\ref{eq:R-Full-System})], and a modulated resonant oscillation followed by an asymptotic contraction  [Figs.~\ref{fig:3-intro-extension}e and \ref{fig:Modulated Resonant with a Subsequent Asymptotic Contraction}; Eqs.~(\ref{eq:R(z)-Fu-Geometric})--(\ref{eq:gamma_{3}(z)})].
With this extension, we expanded  the set of known analytical solutions to the Moore equation from the few classes reported in the literature (Fig.~\ref{fig:Fig-literature}) to an infinite set (in  Fig.~\ref{fig:3-intro-extension} we illustrate some of them), as it enables the treatment of new classes of laws of motion by combining those presented here with each other as well as with previously known classes of trajectories.


\begin{acknowledgments}

The authors thank Edney R. Granhen for useful discussions.
M.F.A. was partially supported by the Fundação Amazônia de
Amparo a Estudos e Pesquisas (FAPESPA), Brazil, through the
PIBIC/FAPESPA program.
M.F.A, O.J.N.N and S.S.C were partially supported by Coordenação de Aperfeiçoamento de Pessoal de Nível Superior (CAPES) - Brazil, Finance Code 001. 
D.T.A was partially supported by 
Conselho Nacional de Desenvolvimento Científico e Tecnológico
(CNPq) - Brazil, under Grant No. 309521/2026-2.
This work was also partially supported by CNPq - Brazil, under Grant No. 408735/2023-6.
\end{acknowledgments}


\appendix
\section{A DETAILED DERIVATION OF EQ. (\ref{eq:tn_Law})}
\label{apendice:recursao_temporal}

Here, we shall detail the algebraic development for the analytical construction of the function $t_{n}(z)$ for the oscillatory law of motion (\ref{eq:Law_motion}). To this end, we employ Eqs. (\ref{t1(z)}) and (\ref{recorrencia-ti}) introduced in Section \ref{Geometric-Method}, considering the law of motion defined in Eq. (\ref{eq:Law_motion}). First, to determine $t_{1}(z)$, we substitute Eq. (\ref{eq:Law_motion}) into (\ref{t1(z)}), which yields 
\begin{equation}
k_{0}t_{1}+\arcsin[\sin\theta\cos(k_{0}t_{1})]-\theta=k_{0}(z-L_{0}),
\end{equation}
where we defined $k_{0}=2\pi/L_{0}$. After some straightforward trigonometric
manipulations, we find
\begin{equation}
k_{0}t_{1}=\arctan\beta_{1}+\pi w_{1},\label{ape:t1(z)-final}
\end{equation}
where
\begin{equation}
\beta_{1}=\frac{\sin(k_{0}z+\theta)-\sin\theta}{\cos(k_{0}z+\theta)},
\end{equation}
and the term $\pi w_{1}$, with $w_{1}\in\mathbb{Z}$, represents the branches of the arctan function, which must be appropriately selected to ensure the continuity of $R(z)$. After further trigonometric manipulations, one can rewrite $\beta_{1}$ as 
\begin{equation}
\beta_{1}=\frac{2\tan(\kappa_{0}z)\left[1-\tan\theta\tan(\kappa_{0}z)\right]}{1-2\tan\theta\tan(\kappa_{0}z)-\tan^{2}(\kappa_{0}z)},
\label{eq:beta1}
\end{equation}
where we also defined $\kappa_{0}=k_{0}/2=\pi/L_{0}.$

Next, we use the recurrence relation [Eq. (\ref{recorrencia-ti})] to obtain the next term $t_{2}(z)$, namely
\begin{equation}
t_{2}+L(t_{2})=2t_{1}-z.
\end{equation}
Analogously to the derivation of $t_{1}$, the expression for $t_{2}$ assumes a similar form, 
\begin{equation}
k_{0}t_{2}=\arctan\beta_{2}+\pi w_{2},
\label{ape:t2_inicial}
\end{equation}
with the replacement $\kappa_{0}z\rightarrow k_{0}t_{1}-\kappa_{0}z$
in the tangent argument of $\beta_{1}$, namely
\begin{equation}
\beta_{2}=\frac{2\tan(k_{0}t_{1}-\kappa_{0}z)[1-\tan\theta\tan(k_{0}t_{1}-\kappa_{0}z)]}{1-2\tan\theta\tan(k_{0}t_{1}-\kappa_{0}z)-\tan^{2}(k_{0}t_{1}-\kappa_{0}z)}.
\end{equation}
Applying the formula for the tangent of a difference in the $\tan(k_{0}t_{1}-\kappa_{0}z)$ term, and then substituting Eq. (\ref{ape:t1(z)-final}) into it, after some algebraic manipulation, it is simplified to
\begin{equation}
\tan(k_{0}t_{1}-\kappa_{0}z)=\frac{\tan(\kappa_{0}z)}{1-2\tan\theta\tan(\kappa_{0}z)}.
\end{equation}
Substituting this result into $\beta_{2}$ and performing further simplifications, 
\begin{equation}
\beta_{2}=\frac{2\tan(\kappa_{0}z)[1-3\tan(\kappa_{0}z)\tan\theta]}{[1-3\tan\theta\tan(\kappa_{0}z)]^{2}-\tan^{2}(\kappa_{0}z)\sec^{2}\theta},
\end{equation}
and finally substituting in Eq. (\ref{ape:t2_inicial}), we obtain the final form of $t_{2}(z)$.

Applying the same procedure recursively for the next terms $t_{3}$, $t_{4}$, \ldots , and identifying the pattern, the general analytical solution for $t_{n}(z)$ is given by Eq. (\ref{eq:tn_Law}).

\section{TREATMENT OF BRANCHES IN EQ. (\ref{eq:tn_Law})}

\label{apendice: Branches in R(z)}

Here, we present the analytical treatment of the arctan function branches
in $t_{n}(z)$ {[}Eq. (\ref{eq:tn_Law}){]}, aiming to ensure the
continuity of $R(z)$, as required. Let us define the auxiliary functions
$N(z)$ and $D(z)$ as the numerator and denominator, respectively,
of $\beta_{n}$ in Eq. (\ref{eq:tn_Law}). Since the arctan function
is restricted to its principal branch in the interval $(-\pi/2,\pi/2)$,
the sign inversion of the denominator $D(z)$ introduces jump discontinuities
in $t_{n}(z)$. In order to eliminate these, we perform continuous
phase unwrapping \cite{Moir-JSC-2021} using the adjustment parameter
$w_{n}$.

This correction is applied sector by sector in a $D$ \emph{versus}
$N$ plane. While the first quadrant ($D>0$ and $N>0$) naturally
remains within the principal branch, the other regions require specific
shifts. Regarding the left half-plane (where $D<0$), when the function
enters the second quadrant, the denominator $D(z)$ becomes negative,
resulting in a jump of $-\pi$ units, which causes the phase to return
to the interval $(-\pi/2,0)$. In the third quadrant, the ratio $N(z)/D(z)$
becomes positive (given $N<0$ and $D<0$), with values returning
to the range $(0,\pi/2)$. In both scenarios, to ensure continuity
by mapping the phase to the intervals $(\pi/2,\pi)$ and $(\pi,3\pi/2)$,
respectively, a $+\pi$ translation is required. This is analytically
implemented by introducing the Heaviside term $\pi\Theta(-D)$. Similarly,
for the transition to the fourth quadrant where $D>0$ and $N<0$,
as the denominator becomes positive again, the natural jump vanishes
and the arctan function falls back into the interval $(-\pi/2,0)$.
Therefore, to ensure the continuity, a jump of $2\pi$ must be included.
Since this condition is exclusive to the region where the denominator
is positive and the numerator is negative, the correction can be implemented
with the additional Heaviside term $2\pi\Theta(D)\Theta(-N)$.

Furthermore, it becomes necessary to include correction terms when
$z$ is an integer multiple of $L_{0}$. This requirement arises from
the periodic nature of the function $\tan(\pi z/L_{0})$, which completes
a cycle of $\pi$ radians every interval $L_{0}$, causing the argument
of the arctan to return to its initial value. However, the treatment
of these periodic discontinuities in $t_{n}(z)$ must be carried out
carefully: the transitions in $t_{n}$ should not be entirely neutralized,
but rather synchronized with the discrete jumps of the function $2n(z)$,
which also occur at integer multiples of $L_{0}$. This mutual compensation
between the dynamics of $t_{n}(z)$ and the increments of $2n(z)$
is the fundamental condition for ensuring the continuity of the function
$R(z)$ across the entire domain. For an overview of how to introduce
these Heaviside functions, it is convenient to decompose the main
function {[}Eq. (\ref{R(z)-Geometric}){]} into the form $R(z)=R_{n}(z)+R_{t_{n}}(z)$,
as follows 
\begin{equation}
R_{n}=2n(z),
\label{eq:Rn_def}
\end{equation}
\begin{equation}
R_{t}(z)=[t_{n}(z)-L(t_{n})]/L_{0},
\label{ape:Rtn_def}
\end{equation}
where 
\begin{equation}
k_{0}t_{n}=\arctan(N/D)+\pi [\Theta(-D)+2\Theta(D)\Theta(-N) ].
\label{ape:tn_full-before}
\end{equation}
\begin{figure}[b]
	\includegraphics[width=1.00\columnwidth]{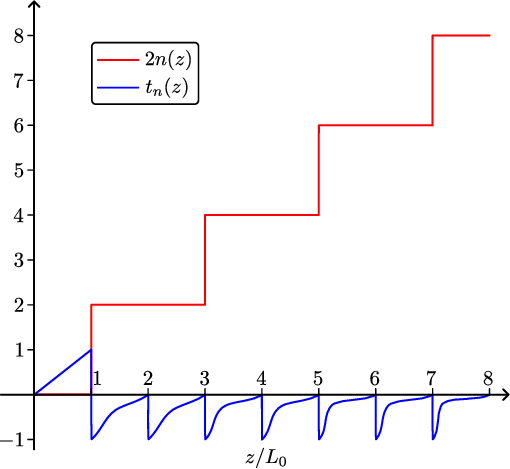}
	\caption{Plots of $t_{n}(z)$ (blue line) and $2n(z)$ (red line). The interval
		$0<z\le L_{0}$ is the static region.}
	\label{Grafic-n(z)-tn(z)} 
\end{figure}
From Fig. \ref{Grafic-n(z)-tn(z)}, it can be observed that the function
$R_{n}(z)$ (red line) exhibits a jump of $2$ units when $z$ equals
an odd multiple of $L_{0}$. On the other hand, the function $R_{t}$
(blue line) drops by $1$ unit at all integer multiples of $L_{0}$.
Consequently, the combination $R(z)=R_{n}(z)+R_{t}(z)$ has a jump
of $1$ when $z$ assumes values equal to odd multiples of $L_{0}$,
and a drop of $1$ unit at even multiples. Taking this into account,
to guarantee the continuity of $R(z)$, a square wave-like correction
($1+\text{sgn}[\cos(\kappa_{0}z)]$) must be included. Consequently,
the final form of the adjustment term $w_{n}(z)$, responsible for
ensuring the continuity of the function $R(z)$, is written as
\begin{equation}
w_{n}=1+\text{sgn}[\cos(\kappa_{0}z)]+\Theta(-D)+2\Theta(D)\Theta(-N).\label{ape:wn(z)}
\end{equation}

\section{DERIVATION OF EQ. (\ref{R-resonant})}
\label{apendice: Trigonometric Manipulations}

The objective of this section is to perform the algebraic manipulations
of the phase function $R(z)$, in order to analytically obtain the
exact solution for the resonant motion. First, we work with the term
$L(t_{n})$ in Eq. (\ref{R(z)-Geometric}) rewriting all trigonometric
dependencies in terms of tangents. We evaluate the argument of $\cos(k_{0}t_{n})$
in Eq. (\ref{eq:Law_motion}) with $t_{n}$ given by Eq. (\ref{eq:tn_Law}).
Using the formula for the cosine of a sum, 
\begin{eqnarray}
\cos(k_{0}t_{n}) & = & \cos(\arctan\beta_{n})\cos(\pi w_{n})\nonumber\\
 &  & -\sin(\arctan\beta_{n})\sin(\pi w_{n}).
\end{eqnarray}
Since $w_{n}$ takes strictly integer values, $\sin(\pi w_{n})=0$
and $\cos(\pi w_{n})=(-1)^{w_{n}}$. Furthermore, analyzing Eq. (\ref{ape:wn(z)}),
except for $\Theta(-D)$ the terms of $w_{n}$ always result in even
integers. Therefore, the parity of $w_{n}$ is entirely determined
by the Heaviside function $\Theta(-D)$, allowing the simplification
$(-1)^{w_{n}}=(-1)^{\Theta(-D)}$ {[}recalling that $N$ and $D$
are the numerator and denominator, respectively, of $\beta_{n}$ in
Eq. (\ref{eq:tn_Law}){]}. Given that the principal branch of $\arctan x$
restricts the angle to the interval $(-\pi/2,\pi/2)$, the cosine
remains positive, such that $\cos(\arctan x)=1/\sqrt{1+x^{2}}$. Applying
this identity, we find 
\begin{equation}
\cos(k_{0}t_{n})=\frac{(-1)^{\Theta(-D)}|D|}{\sqrt{N^{2}+D^{2}}}=\frac{D}{\sqrt{N^{2}+D^{2}}}.\label{ape:cos(tn)}
\end{equation}

Applying Eq. (\ref{ape:cos(tn)}) into (\ref{eq:Law_motion}) and
using the identity $\sin\theta=\tan\theta/\sqrt{1+\tan^{2}\theta}$,
we obtain 
\begin{equation}
k_{0}L(t_{n})=2\pi-\theta+\arcsin\left(\frac{vD}{\sqrt{1+v^{2}}\sqrt{N^{2}+D^{2}}}\right),\label{eq:k0Ltn}
\end{equation}
where we defined $v=\tan\theta$. To proceed, we utilize the relation
\cite{Gradstein1982tables}
\begin{equation}
\arcsin x=\arctan\frac{x}{\sqrt{1-x^{2}}},
\end{equation}
which is valid for $x^{2}<1$. After simplification, Eq. (\ref{eq:k0Ltn})
becomes
\begin{equation}
k_{0}L(t_{n})=2\pi-\theta+\arctan\frac{vD}{T},
\label{ape:L(tn(z))}
\end{equation}
where we defined $u=\tan(\kappa_{0}z)$ and $T=[1-(2n-1)uv]^{2}+u^{2}(1+v^{2})$.

We now evaluate the difference $t_{n}-L(t_{n})$. From Eqs. (\ref{ape:L(tn(z))}) and (\ref{eq:tn_Law}),
we get
\begin{eqnarray}
k_{0}[t_{n}-L(t_{n})]+2\pi & = & \arctan\frac{N}{D}+\theta\nonumber \\
 &  & -\arctan\frac{vD}{T}+\pi w_{n}.
 \label{ape:tn-L(tn)}
\end{eqnarray}
We shall apply identity formulas in these combinations of arctan functions. First, we analyze the sum $\arctan(N/D)+\theta$.
Using the property \cite{Gradstein1982tables}
\begin{equation}
\arctan x\pm\arctan y=\arctan\frac{x\pm y}{1\mp xy}+s\pi,
\label{eq:arctan_sum}
\end{equation}
with
\begin{equation}
s=\begin{cases}
\mathrm{sgn}\,x, & \pm xy>1\\[6pt]
0, & \pm xy<1
\end{cases},
\end{equation}
and after the appropriate algebraic manipulations, we obtain
\begin{equation}
\arctan\frac{N}{D}+\theta=\arctan\frac{N+vD}{D-vN}+\ell_{1}\pi.
\label{ape:gamma}
\end{equation}
At this stage, the branches are controlled by the Heaviside function
$\ell_{1}=\Theta(vN/D-1)$. Substituting Eq. (\ref{ape:gamma}) into
Eq. (\ref{ape:tn-L(tn)}), we get
\begin{eqnarray}
k_{0}[t_{n}-L(t_{n})]+2\pi & = & \arctan\frac{N+vD}{D-vN}-\arctan\frac{vD}{T}\nonumber \\
 &  & +\ell_{1}\pi+w_{n}\pi.
 \label{ape:tn-L(tn)-1}
\end{eqnarray}
We apply identity (\ref{eq:arctan_sum}) once again for the subtraction
of arctan terms in Eq. (\ref{ape:tn-L(tn)-1}). After some simplifications,
the expression reduces to 
\begin{eqnarray}
k_{0}[t_{n}-L(t_{n})]+2\pi & = & \arctan\frac{2u(1-2nvu)}{(1-2nvu)^{2}-u^{2}}\nonumber \\
 &  & +\ell_{1}\pi+\ell_{2}\pi+w_{n}\pi,
 \label{eq:tn_Ltn_flat}
\end{eqnarray}
with $\ell_{2}=\Theta(-xy-1)\mathrm{sgn}\,x$, where $x$ and $y$
are the arguments of the first and second arctan terms in Eq. (\ref{ape:tn-L(tn)-1}),
respectively.

By using the identity
\begin{equation}
\arctan\frac{2x}{1-x^{2}}=2\arctan x+s\pi,
\end{equation}
where
\begin{equation}
s=\begin{cases}
\mathrm{sgn}\,x, & |x|>1\\[6pt]
0, & |x|<1
\end{cases},
\end{equation}
in Eq. (\ref{eq:tn_Ltn_flat}) with $x$ being identified as $x=u/(1-2nvu)$,
we can rewrite the term as $2\arctan x+\ell_{3}\pi$, where the new
branch control function is defined as $\ell_{3}=-\mathrm{sgn}\,x\,\Theta(|x|-1)$.
After simplifications, we have 
\begin{eqnarray}
k_{0}[t_{n}-L(t_{n})]+2\pi & = & 2\arctan\frac{u}{1-2nvu}\nonumber \\
 &  & +(\ell_{1}+\ell_{2}+\ell_{3}+w_{n})\pi.
\end{eqnarray}
Finally, we apply the property \cite{Gradstein1982tables}
\begin{equation}
\arctan x=\ell_{4}\pi/2-\arctan(1/x),
\end{equation}
where $\ell_{4}=\mathrm{sgn}\,x$. Therefore, rearranging the terms
we obtain
\begin{eqnarray}
\frac{t_{n}-L(t_{n})}{L_{0}} & = & -\frac{1}{\pi}\arctan[\cot(\kappa_{0}z)-k_{0}n\epsilon]\nonumber \\
 &  & -1+\frac{\ell_{1}+\ell_{2}+\ell_{3}+\ell_{4}+w_{n}}{2}.
 \label{ape:tn-Ltn_final}
\end{eqnarray}

Analyzing the combination of the step functions responsible for controlling
the branches of the $\arctan$ function, they always result in $1$
for any configuration of their arguments, namely
\begin{equation}
\Theta(-D)+2\Theta(D)\Theta(-N)+\ell_{1}+\ell_{2}+\ell_{3}+\ell_{4}=1.
\end{equation}
In this way, Eq. (\ref{ape:tn-Ltn_final}) simplifies to 
\begin{eqnarray}
\frac{t_{n}-L(t_{n})}{L_{0}} & = & \frac{1}{2}-\frac{1}{\pi}\arctan[\cot(\kappa_{0}z)-k_{0}n\epsilon]\nonumber \\
 &  & +\sum^{\infty}_{j=1}(-1)^{j}\Theta(z-jL_{0}).
 \label{ape:tn_final_final}
\end{eqnarray}
Finally, by substituting Eq.~(\ref{ape:tn_final_final}) into (\ref{R(z)-Geometric}),
we obtain the final expression for the function $R(z)$ given by
Eq. (\ref{R-resonant}).


%

\end{document}